\documentclass[twoside,twocolumn]{article}

\usepackage{jzusc}

\usepackage{graphicx}

\begin{document}

\title{Interpolation of a spline developable surface between a 
  curve and two rulings}
\author[$\ddagger$1]{A. Cant\'on and L. Fern\'andez-Jambrina}
\affil[1]{ETSI Navales, Universidad Polit\'ecnica de Madrid, 
Arco de la Victoria 4, 28040-Madrid, Spain}
\authmark{ }

\corremailB{leonardo.fernandez@upm.es} 
\corremailA{alicia.canton@upm.es}
\emailmark{ }





%
\abstract{
In this paper we address the problem of interpolating a spline
developable patch bounded by a given spline curve and the first and
the last rulings of the developable surface. In order to complete the
boundary of the patch a second spline curve is to be given. Up to now 
this interpolation problem could be solved, but without the 
possibility of choosing both endpoints for the rulings. We circumvent 
such difficulty here by resorting to degree elevation of the 
developable surface. This is useful not only to solve this problem, 
but also other problems dealing with triangular developable patches.}
    
\keywords{
Developable surfaces, Spline surfaces, blossoms.
}


\shortauthor{Cant\'on et al.}

\doi{10.1631/jzus.C1000000}
\code{A}
\clc{}

\publishyear{2011}
\vol{12}
\issue{}
\pagestart{1}
\pageend{5}


\articleType{}

\dateinfo{}

\maketitle
\section{Introduction}
\label{intro}

Developable surfaces have been used extensively in industry for
modelling sheets of steel.  These surfaces are plane patches that have
been curved by isometric transformations, preserving lengths of
curves, angles and areas. They mimic the properties of thin steel 
plates that are transformed by cutting, rolling or folding, but not 
by stretching or application of heat, which would raise manufacturing 
costs. 

Their inclusion in the NURBS formalism, however, has not been easy. 
The condition of developability is a non-linear differential equation 
which translates into non-linear equations for the vertices of the 
control net of the surface.

To our knowledge the first reference to NURBS developable surfaces 
arises in technical reports at General Motors (\cite{mancewicz,frey}). One approach has been solving the developability 
condition for low degrees (\cite{lang,sequin,strip}).  

Another approach to developable surfaces consists in resorting to 
projective dual geometry. In this geometry ``points'' are planes and 
``planes'' are points and this is useful to solve the developability 
condition (\cite{ravani,pottmann-farin,hu}).

One can also construct surfaces which are approximately developable
instead (\cite{chalfant,wallner,leopoldseder,peternell,liu,liu1}).  A nice
review may be found in  \cite{computational-line}.  Applications
to ship hull design may be found in 
\cite{kilgore,arribas,leonardo-hull}.

A large family of B\'ezier developable surfaces was obtained in
\cite{aumann,aumann1} defining affine transformations between
cells of the control net. This result has been extended to spline 
(\cite{leonardo-developable}) and B\'ezier triangular 
(\cite{leonardo-triangle}) 
developable patches. A characterisation of B\'ezier ruled surfaces is 
found in \cite{juhasz-ruled}.

In this paper we make use of the latter constructions 
to find solutions to interpolation problems with developable 
surfaces. For instance, in \cite{leonardo-developable}, we were 
able to draw a developable surface through a given boundary curve and 
two rulings, but we could not choose both endpoints for these 
rulings. We would like to solve such an issue and also apply the 
solution to new problems.

Following \cite{leonardo-developable}, we first review in Section
\ref{develop} the main features, definitions and the classification of
developable surfaces, whereas in Section~\ref{curves} we deal with the
formalism of B-spline curves.  In Section~\ref{splines} we review the
construction of spline developable surfaces grounded on linear
relations between vertices of the B-spline net, that was given in
\cite{leonardo-developable}.  In Section~\ref{interpol} we use that
construction to provide solutions to an interpolation problem between
a spline curve and two rulings as in \cite{leonardo-developable}.
Finally, in Section~\ref{elevate} we use degree elevation to provide 
our new solution to the problem of interpolating a developable patch 
between a spline curve and segments of the rulings at both ends.  
This problem could
not be solved with just our previous results.  This solution is
extended to triangular patches in Section~\ref{triangles}.  A final
section of conclusions is included at the end of the paper.

\section{Developable surfaces}
\label{develop}
A \emph{ruled surface} patch fills the space between two parametrised 
curves $c(u)$, $d(u)$,
\begin{equation}\label{ruled}
b(u,v)=(1-v)c(u)+vd(u),\ u\in[a,b],\end{equation} for $v\in[0,1]$, 
by linking with segments, named \emph{rulings}, the points on both curves
with the same parameter $u$.

In general, the tangent plane to the ruled surface on a 
ruling is different for each point on the segment. \emph{Developable 
surfaces} are the subcase of ruled surfaces for which the tangent 
plane is constant along each ruling (\cite{struik,postnikov}). 

Let us compute a normal vector at each point of a ruled surface with
the derivatives of the parametrisation in Eq.~\ref{ruled},
\[b_{u}(u,v)=(1-v)c'(u)+vd'(u),\ b_{v}(u,v)=d(u)-c(u),\]
\[\left(b_{u}\times b_{v}\right)(u,v)=\left((1-v)c'(u)+vd'(u)\right)\times\left(
d(u)-c(u)\right),\]
which is linear in the parameter $v$. If we calculate it on both ends of the rulings,
\[\left(b_{u}\times b_{v}\right)(u,0)=c'(u)\times\left(
d(u)-c(u)\right),\]\[
\left(b_{u}\times b_{v}\right)(u,1)=d'(u)\times\left(
d(u)-c(u)\right),\]
we learn that the three vectors $c'(u)$, $d'(u)$, $d(u)-c(u)$ are to 
be coplanary in order to have a constant tangent plane along each 
ruling  of the surface.

\noindent \textbf{Proposition:} A ruled surface parametrised as in 
Eq.~\ref{ruled} is
developable if and only if the vector $\mathbf{w}(u)=d(u)-c(u)$,
linking the points $d(u)$, $c(u)$, and the velocities $c'(u)$, 
$d'(u)$ of the curves at
these points are coplanary for every value of $u$.

%
%
%
%
%
%

\section{B-spline curves}
\label{curves}
In this section we review the formalism of B-spline curves and their 
main properties in order to fix the notation, which follows closely 
the one in \cite{farin}. 
We may define a B-spline curve $c(u)$ of degree $n$ and $N$ pieces on 
an interval $[u_{n-1},u_{n+N-1}]$, so that the $I$-th piece of the 
curve is defined on an interval $[u_{n+I-2},u_{n+I-1}]$. For this we 
require an ordered list of values of the parameter $u$, which are named 
\emph{knots}, $\{u_{0},\ldots,u_{2n+N-2}\}$. The actual knots 
defining the intervals for each piece are the \emph{inner} knots 
$\{u_{n-1},\ldots ,u_{n+N-1}\}$ whereas the 
knots $\{u_{0},\ldots,u_{n-2}\}$ at the beginning of the 
list (usually taken to be equal to $u_{n}$) and $\{u_{n+N},\ldots, u_{2n+N-2}\}$  at the end
(usually taken to be equal to $u_{n+N-1}$) are \emph{auxiliary}. 

Points on B-spline curves can be computed using the De
Boor's algorithm, $c(u)=c^{n)}_0(u)$, consisting on linear interpolations between consecutive
vertices. For a curve of just one piece:
\begin{eqnarray}\label{deboor1}
c^{r)}_i(u)&:=&
\frac{u_{i+n}-u}{u_{i+n}-u_{i+r-1}}c^{r-1)}_{i}(u)\nonumber\\&+&
\frac{u-u_{i+r-1}}{u_{i+n}-u_{i+r-1}}c^{r-1)}_{i+1}(u),
\end{eqnarray}
for $i=0,\ldots,n-r$, $r=1,\ldots,n$.

A useful construction, named \emph{polarisation} or \emph{blossom} of 
the parametrisation of the curve, consists of interpolating in each 
step with a different value $v_{i}$ of the parameter $u$, 
$c[v_1,\ldots,v_n]:=c^{n)}_0[v_1,\ldots,v_{n}]$,
\begin{eqnarray}\label{deboorb}
\!\!\!\!c^{r)}_i[v_1,\ldots,v_r]\!\!\!\!\!&:=&\!\!\!\!\!
\frac{u_{i+n}-v_r}{u_{i+n}-u_{i+r-1}}c^{r-1)}_{i}[v_1,\ldots,v_{r-1}]
\nonumber\\\!\!\!\!\!& +&\!\!\!\!\!
 \frac{v_r-u_{i+r-1}}{u_{i+n}-u_{i+r-1}}c^{r-1)}_{i+1}[v_1,\ldots,v_{r-1}].
\end{eqnarray}

With this notation,
$u^{<i>}=\underbrace{u,\ldots,u}_{i\ \mathrm{times}}$, 
we have that $c(u)=c[u^{<n>}]$.  Vertices are recovered from the 
polarisation as $c_{i}=c[u_{i},\ldots,u_{i+n-1}]$.

These expressions are valid for B-spline curves with an arbitrary number of 
pieces, replacing the interval $[u_{n-1},u_{n}]$ of the first 
piece by the interval of the piece under consideration. 

We may summarise some properties of the De Boor algorithm and the 
polarisation which are 
relevant for our purposes:
\begin{enumerate}
\item The velocity of the curve is 
\begin{eqnarray}\label{velo}\!\!\!\!\!\!\!\!\!c'(u)\!\!\!\!\!&=&\!\!\!\!\!
    \displaystyle\frac{n}{u_{n}-u_{n-1}}\left(
c_1^{n-1)}(u)-c_0^{n-1)}(u)\right)\nonumber\\
\!\!\!\!\!&=&\!\!\!\!\!\frac{n\left(
c[u^{<n-1>},u_{n}]-
c[u^{<n-1>},u_{n-1}]\right)}{u_{n}-u_{n-1}}.\end{eqnarray}
\item The polarisation $c[v_1,\ldots,v_n]$ of the spline curve $c(u)$, is multiaffine and 
symmetric. That is, if $\lambda+\mu=1$,
\begin{eqnarray*}\label{maffine}\!\!\!\!\!\!\!\!\!\!\!\!
    c[\lambda v_1+\mu \tilde v_{1},\ldots,v_n]=\lambda 
c[v_1,\ldots,v_n]+\mu c[\tilde v_1,\ldots,v_n].\end{eqnarray*}
\end{enumerate}

Finally, we review two operations with B-spline curves which we shall 
need later on:

\noindent \textbf{Insertion of knots:} Given a B-spline curve of
degree $n$ with vertices $\{c_{0},\ldots,c_{L}\}$ and knots
$\{u_{0},\ldots,u_{K}\}$, we can split into two the piece
corresponding to the interval $[u_{I},u_{I+1}]$ by inserting a new
knot $\tilde u$, $u_{I}<\tilde u<u_{I+1}$.  The new list of knots is
then obviously $\{\tilde u_{0},\ldots,\tilde u_{K+1}\}$,
\[\tilde u_{i}=u_{i},\ i=0,\ldots,I,\ \tilde u_{I+1}=\tilde 
u,\ \tilde u_{i}=u_{i-1},\] for $i=I+2,\ldots,K+1$, 
and, since the curve has not changed, the blossom provides the new 
sequence of vertices $\{\tilde c_{0},\ldots,\tilde c_{L+1}\}$,
\[\tilde c_{i}=c[\tilde u_{i},\ldots,\tilde u_{i+n-1}],\qquad 
i=0,\ldots,L+1.\]

\noindent \textbf{Degree elevation:} Formally we may express a B-spline curve 
$c(u)$ of degree $n$ as a curve of degree $n+1$. The blossom $c^1$ of the degree-elevated  
curve is related to the original one in a simple form (\cite{farin}),
\begin{eqnarray}\label{elevboor}
c^1[v_{1},\ldots,v_{n+1}]\!=\!\frac{\displaystyle\sum_{i=1}^{n+1}c[v_{1},\ldots,
v_{i-1},v_{i+1},\ldots,v_{n+1}]}{n+1}\!\!\!
\end{eqnarray}
and in the list of knots $\{u_{0},\ldots, u_{K}\}$ the multiplicity of
inner knots, from $u_{n-1}$ to $u_{n+N-1}$, is increased by one,
without modifying the auxiliary knots.  

\section{Spline developable surfaces}
\label{splines}
The developability condition in Proposition~1 may be readily now adapted 
to spline curves (\cite{leonardo-developable}).

To start, let us consider two B-spline curves of degree $n$ and one
segment over a common list of knots $\{u_{0},\ldots,u_{2n-1}\}$, 
defined on the interval $[u_{n-1},u_{n}]$. 
Their respective B-spline polygons are $\{c_{0},\ldots,c_{n}\}$,
$\{d_{0},\ldots,d_{n}\}$. 

\begin{figure}[h]\begin{center}
\includegraphics[height=0.2\textheight]{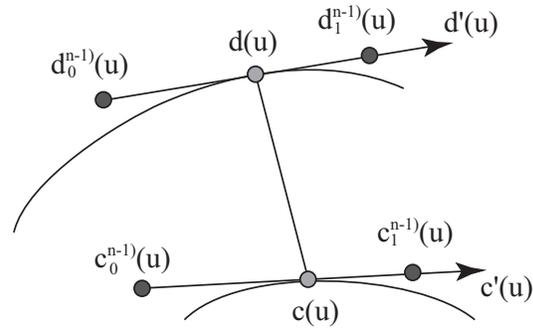}
\caption{Characterisation of developable surfaces}
\label{plane}\end{center}
\end{figure}

We may draw a simple conclusion using the De Boor algorithm. Using 
Eq.~\ref{velo} and the last iteration of Eq.~\ref{deboor1}, 
 it is clear that the vectors $c'(u)$, $d'(u)$, $d(u)-c(u)$ are 
coplanary if and only if the four points $c_0^{n-1)}(u)$, 
$c_1^{n-1)}(u)$, $d_0^{n-1)}(u)$, $d_1^{n-1)}(u)$ are coplanary (see 
Figure~\ref{plane}).

The developability condition is then equivalent to the possibility of 
writing one of the points as a barycentric combination of the other 
ones. For instance,
\begin{eqnarray*}d_1^{n-1)}(u)&=&\mu_{0}(u)d_0^{n-1)}(u)+
\lambda_{0}(u)c_0^{n-1)}(u)\\&+&\lambda_{1}(u)c_1^{n-1)}(u),
\end{eqnarray*}
with coefficients $\lambda_{0}(u)$, $\lambda_{1}(u)$, $\mu_{0}(u)=
1-\lambda_{0}(u)-\lambda_{1}(u)$.

We may rewrite this combination in another form, separating the terms 
related to each curve, also in a barycentric fashion,
\begin{eqnarray}\label{barycentric}&&(1-\Lambda(u))c_0^{n-1)}(u)+\Lambda(u)c_1^{n-1)}(u)
\nonumber\\&&=
(1-M(u))d_0^{n-1)}(u)+M(u)d_1^{n-1)}(u),\end{eqnarray}
\[\Lambda(u)=\frac{\lambda_{1}(u)}{\lambda_{0}(u)+\lambda_{1}(u)},\quad
M(u)=\frac{1}{\lambda_{0}(u)+\lambda_{1}(u)},\]
which just excludes the case of parallel vectors
$d_{1}^{n-1)}(u)-d_{0}^{n-1)}(u)$, $c_{1}^{n-1)}(u)-c_{0}^{n-1)}(u)$, 
which corresponds to a cone. In this sense we use the word 
\emph{generic}, since the following results will be valid for all 
developable surfaces, but for this type of cone.

Using blossoms and taking into account that 
these are multiaffine Eq.~(\ref{maffine}),
\begin{eqnarray*}
    &&(1-\Lambda(u))c_0^{n-1)}(u)+\Lambda(u)c_1^{n-1)}(u)\nonumber
\\&&=
(1-\Lambda(u))c[u^{<n-1>},u_{n-1}]+\Lambda(u)c[u^{<n-1>},u_{n}]\\&&=
c\left[u^{<n-1>},(1-\Lambda(u))u_{n-1}+\Lambda(u)u_{n}\right],\end{eqnarray*}
the coplanarity condition (Eq.~\ref{barycentric}) may be written in a 
more compact expression,
\begin{equation}\label{condi}c[u^{<n-1>},\Lambda^*(u)]=
d[u^{<n-1>},M^{*}(u)],\end{equation}
\[\Lambda^*(u)=(1-\Lambda(u))u_{n-1}+\Lambda(u) u_{n},\]\[
M^{*}(u)=(1-M(u))u_{n-1}+M(u) u_{n}.\]


This expression is valid for B-spline curves with arbitrary number of 
pieces, replacing the interval $[u_{n-1},u_{n}]$ of the first 
piece by the interval of the piece under consideration. 

The higher the degree of $\Lambda^{*}(u)$, $M^{*}(u)$, the larger the 
number of conditions imposed by Eq.~\ref{condi}. Hence,  
we restrict now to the case with constant $\Lambda^*$, $M^*$, which
produces the families of developable surfaces in
\cite{aumann,leonardo-developable}.  In this case expressions on both
sides of Eq.~\ref{condi} may be viewed as parametrisations of curves of
degree $n-1$ and therefore this condition is equivalent to the same
one for their blossoms, since a blossom is uniquely determined by its 
parametrisation: 
\begin{theorem} Two B-spline curves of degree $n$ and $N$ pieces with the same list of knots
 $\{u_{0},\ldots,u_{K}\}$ define a  developable surface on the 
 interval $[u_{n-1},u_{n+N-1}]$ if 
their blossoms are related by 
\[c[v_{1},\ldots,v_{n-1},\Lambda^*]=d[v_{1},\ldots,v_{n-1},M^*],\]
for some values $\Lambda^*$, $M^*$.
\end{theorem}

We may obtain relations between the B-spline polygons of both 
curves by applying the previous expression to lists of correlative knots,
$\{u_{i+1},\ldots, u_{i+n-1}\}$, taking into account that blossoms 
are multiaffine,
\begin{eqnarray*}&&c[u_{i+1},\ldots, 
u_{i+n-1},\Lambda^*]\nonumber\\&&=
c\left[u_{i+1},\ldots, 
u_{i+n-1},\frac{u_{i+n}-\Lambda^*}{u_{i+n}-u_{i}}u_{i}+\right.\\&&\left.
\frac{\Lambda^*-u_{i}}{u_{i+n}-u_{i}}u_{i+n}\right]\\&&=
\frac{u_{i+n}-\Lambda^*}{u_{i+n}-u_{i}}c\left[u_{i},\ldots, 
u_{i+n-1}\right]\\&&+\frac{\Lambda^*-u_{i}}{u_{i+n}-u_{i}}
c\left[u_{i+1},\ldots, u_{i+n}\right]\\&&=
\frac{u_{i+n}-\Lambda^*}{u_{i+n}-u_{i}}c_{i}+\frac{\Lambda^*-u_{i}}{u_{i+n}-u_{i}}
c_{i+1},\end{eqnarray*}
since $c_{i}=c[u_{i},\ldots,u_{i+n-1}]$.

\noindent \textbf{Corollary 1:} Two B-spline curves of degree $n$ with the same list of knots
 $\{u_{0},\ldots,u_{K}\}$ and B-spline polygons 
 $\{c_{0},\ldots,c_{L}\}$, $\{d_{0},\ldots,d_{L}\}$ define a  developable surface 
if the cells of the B-spline net of the surface are plane and their 
vertices are related by
\begin{eqnarray}&&(u_{i+n}-\Lambda^*)
    c_{i}+(\Lambda^*-u_{i}) c_{i+1}\nonumber\\&&=
    (u_{i+n}-M^*)
    d_{i}+(M^*-u_{i})d_{i+1},\label{control}\end{eqnarray} for
some values $\Lambda^*$, $M^*$ and $i=0,\ldots,L-1$.

This family of spline developable surfaces has the advantage of 
being defined by linear relations between vertices, in spite of the 
non-linearity of the condition of null gaussian curvature.

The data for this construction are the B-spline polygon 
$\{c_{0},\ldots,c_{L}\}$, the list of knots $\{u_{0},\ldots,u_{K}\}$ and, for instance, the first plane cell of the 
net, given by either $d_{0}$, $d_{1}$ or $d_{0}$ and the parameters 
$\Lambda^*$, $M^*$. 

Since this construction is based on blossoms of curves, it is compatible 
with algorithms for B-spline curves, grounded on blossoms, such as, 
for instance, the knot insertion algorithm for subdivision of 
B-spline curves. That is, if we split into two pieces the interval 
$[u_{I},u_{I+1}]$ by inclusion of a new knot $\tilde u$, so that the 
new list is $\{u_{0},\ldots,u_{I},\tilde u,u_{I+1},\ldots,u_{K}\}$ 
and we compute the new B-spline polygons $\{\tilde 
c_{0},\ldots,\tilde c_{L+1}\}$ and $\{\tilde d_{0},\ldots,\tilde 
d_{L+1}\}$, these new vertices satisfy Eq.~\ref{control}.

\begin{figure}[h]\begin{center}
\includegraphics[height=0.2\textheight]{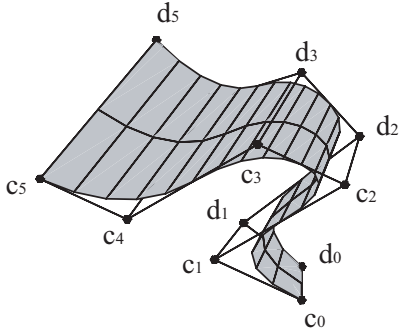}
\caption{Developable B-spline surface of 4 pieces of degree 2}
\label{develo}\end{center}
\end{figure}
\begin{figure}[h]\begin{center}
\includegraphics[height=0.2\textheight]{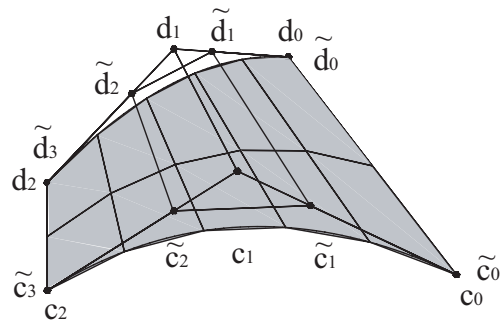}
\caption{Degree-elevated developable surface of one piece of degree 2}
\label{incompatible}\end{center}
\end{figure}

However, this construction is not compatible with degree elevation of
B-spline curves. The degree-elevated B-spline developable surface
through two B-spline curves does not coincide with the B-spline
developable surface through the corresponding degree-elevated curves. 
See, for instance, in Figure~\ref{incompatible} a developable surface 
and the control polygons of the degree-elevated boundary curves 
(denoted by tildes): the central cell of the degree-elevated surface 
is not even planar.

We show it explictly with a simple example: 

\begin{example} Find a developable surface patch of degree two 
and just one piece, bounded by two curves, $c(u)$ and $d(u)$, with 
polygons,
\[c_{0}=(0, 0, 0),\ c_{1}=(3, 3, 0),\ c_{2}=(4, 3, 0);\]\[ d_{0}=(0, 0, 2),\ 
d_{1}=(2, 2, 3),
\]
and knots $\{0,0,1,1\}$. \end{example} 

From Eq.~\ref{control} applied to the first cell of the B-spline net, $i=0$, 
\[(u_{2}-\Lambda^*)c_{0}+(\Lambda^*-u_{0}) 
c_{1}=(u_{2}-M^*)d_{0}+(M^*-u_{0})d_{1},\]
with $n=2$, $u_{0}=0$, $u_{2}=1$, we get 
\begin{eqnarray*}(1-\Lambda^{*})(0, 0, 0)+\Lambda^{*}(3, 3, 0)&=&
    (1-M^{*})(0, 0, 2)\\&+&M^{*}(2, 2, 3),\end{eqnarray*}
and hence $\Lambda^{*}=-4/3$ and $M^{*}=-2$.

We lack the vertex $d_{2}$, but for the second cell of the net,
\[(u_{3}-\Lambda^*)c_{1}+(\Lambda^*-u_{1}) 
c_{2}=(u_3-M^*)d_{1}+(M^*-u_{1})d_{2},\]
\[\frac{7}{3}(3, 3, 0)-\frac{4}{3}(4, 3, 0)=
3(2,2,3)-2d_{2},\]
we conclude $d_{2}=(13/6, 3/2, 9/2)$.

If we formally elevate the degree of both curves to three, the list 
of knots extends to $\{0,0,0,1,1,1\}$ and the new polygons 
obtained with Eq.~\ref{elevboor},
\begin{eqnarray*}
\tilde c_{0}&=&\tilde c[0,0,0]=c[0,0]=c_{0}=(0,0,0)
\\
\tilde c_{1}&=&\tilde c[0,0,1]=\frac{c[0,0]+2c[0,1]}{3}=
\frac{c_{0}+2c_{1}}{3}\\&=&(2,2,0)
\\
\tilde c_{2}&=&\tilde c[0,1,1]=\frac{2c[0,1]+c[1,1]}{3}=
\frac{2c_{1}+c_{2}}{3}\\&=&(10/3,3,0)
\\
\tilde c_{3}&=&\tilde c[1,1,1]=c[1,1]=c_{2}=(4,3,0)
\end{eqnarray*}
\begin{eqnarray*}
\tilde d_{0}&=&\tilde d[0,0,0]=d[0,0]=d_{0}=(0,0,2)
\\
\tilde d_{1}&=&\tilde d[0,0,1]=\frac{d[0,0]+2d[0,1]}{3}=
\frac{d_{0}+2d_{1}}{3}\\&=&(4/3, 4/3, 8/3)
\\
\tilde d_{2}&=&\tilde d[0,1,1]=\frac{2d[0,1]+d[1,1]}{3}=
\frac{2d_{1}+d_{2}}{3}\\&=&(37/18, 11/6, 7/2)
\\
\tilde d_{3}&=&\tilde d[1,1,1]=d[1,1]=d_{2}=(13/6, 3/2, 9/2)
\end{eqnarray*}
correspond to a developable surface with non constant 
$\Lambda^*(u)=-2-u/2$, $M^{*}(u)=-3-u/2$ and it is easy to check that 
the four points that form the second cell, $\tilde c_{1}$, $\tilde 
c_{2}$, $\tilde d_{1}$, $\tilde d_{2}$ do not lie on a plane.

This feature, however, will be shown to be useful for solving interpolation
problems, as it will be apparent in the following sections. 

\section{Interpolation of B-spline developable surfaces}
\label{interpol}
Let us consider the following interpolation problem:

\noindent \textbf{Problem 1:} Given a spline curve $c(u)$ of degree $n$, $N$ 
pieces,  B-spline polygon $\{c_{0},\ldots,c_{L}\}$ and list of knots 
    $\{u_{0},\ldots,u_{K}\}$, $u\in[a,b]$, $a=u_{n-1}$, 
    $b=u_{n+N-1}$, and two
straight lines $l_{a}$ and $l_{b}$ through the endpoints of $c(u)$ 
with respective director vectors $\mathbf{v}$, $\mathbf{w}$, find
a developable surface $b(u,v)$ such that $c(u,0)=c(u)$ and
$l_{a}$ and $l_{b}$ are the first and last rulings of the
surface, that is, $l_{a}:\ c(a,v)$, $l_{b}:\ c(b,v)$.

The special case of B\'ezier curves of degree $n$ was solved
by  \cite{aumann}, making use of his family of developable
surfaces.  His solution is extended to spline curves 
in \cite{leonardo-developable}, solving the 
recursion in Eq.~\ref{control} for the B-spline net. We review here 
this construction in order to extend it to solve new problems in next 
sections. 

We focus on the general case of crossing rulings $l_{a}$ and $l_{b}$,
since the particular cases of parallel or intersecting rulings may be
solved in a simpler fashion resorting to cylinders and cones
respectively.

As in \cite{leonardo-developable}, the last ruling of the developable 
surface can be written in terms of 
the B-spline net of the curve $c(u)$, the list of knots and the 
coefficients $\Lambda^*$, $M^*$,
\begin{eqnarray}\label{recursive1}\!\!\!\!\!\!\!\!\!
    d_{L}-c_{L}\!\!\!\!\!\!&=&\!\!\!\!\!\!\prod _{i=0}^{L-1}\frac
    {M^*-u_{i+n}}{M^*-u_{i}}(d_{0}-c_{0})\nonumber\\
\!\!\!\!\!\!&+&\!\!\!\!\!\!
    \frac{\Lambda^{*}-M^*}{M^*-u_{L-1}}\big(c_{L}-a(M^*) \big),\nonumber\\
\!\!\!\!\!\!\!\!\!    a(M^*)\!\!\!\!\!\!&=&\!\!\!\!\!\!\frac{ M^*-u_{L-1}}{M^*-u_{0} }
    \prod _{i=1}^
	{L-1}{\frac {M^*-u_{i+n}}{M^*-u_{i}}} c_{0} \nonumber\\
\!\!\!\!\!\!&+& \!\!\!\!\!\!\sum _
    {i=1}^{L-1}\frac{ u_{i+n}-u_{i-1} }{ M^*-u_{i-1}}\!\!
    \left(\prod _{j=i}^{L-2}{\frac
    {M^*-u_{n+j+1}}{M^*-u_{j}}}\right)\!\!
      c_{i}.
\end{eqnarray}

From this expression we learn that the vectors along the first and
last rulings, $d_{0}-c_{0}=\sigma\mathbf{v}$,
$d_{L}-c_{L}=\tau\mathbf{w}$, and the vector, $c_{L}-a(M^*)$ have to
be linearly dependent and this will happen for any solution $M^*_{0}$ of 
the algebraic equation
\begin{equation}\label{cramer}\det(a(M^{*})-c_{L},\mathbf{v}, 
    \mathbf{w})=0.\end{equation}

This allows us to write the linear combination in terms of a basis 
$\{\mathbf{v},\mathbf{w},\mathbf{n}\}$, $\mathbf{n}=\mathbf{v}\times 
\mathbf{w}$,
\[a(M^{*}_{0})=c_{L}+\alpha\mathbf{v}+\beta\mathbf{w}+ 0 \mathbf{n},\]
where the coefficients are readily obtained by Cramer's rule,
\[\alpha=\frac{\det(a(M^{*}_{0})-c_{L},\mathbf{w}, \mathbf{n})}
{\det(\mathbf{v},\mathbf{w}, \mathbf{n})},\]\[
\beta=\frac{\det(\mathbf{v},a(M^{*}_{0})-c_{L}, \mathbf{n})}
{\det(\mathbf{v},\mathbf{w}, \mathbf{n})}.\]

Since $M^*$ is fixed by the coplanarity condition in Eq.~\ref{cramer}, 
if we wish, we can modify the length of the rulings through either $\sigma$ or 
$\tau$ just with the parameter 
$\Lambda^{*}$, which remains free so far,
\begin{eqnarray}\label{sigmatau}
\sigma&=&\alpha\frac{\Lambda^{*}-M^*_{0}}
{M^*_{0}-u_{L-1}}  \prod _{i=0}^{L-1}\frac{M^*_{0}-u_{i}}
{M^*_{0}-u_{i+n}},\nonumber\\\tau&=&\beta\frac{M^{*}_{0}-\Lambda^*}
{M^*_{0}-u_{L-1}}.
    \end{eqnarray}
    
\begin{figure}[h]\begin{center}
\includegraphics[height=0.2\textheight]{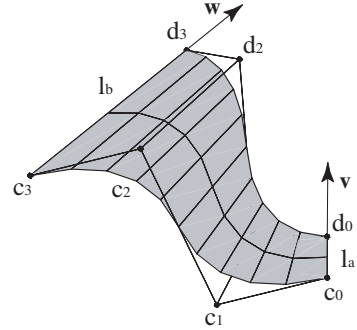}
\caption{Developable surface of degree 2 and 2 pieces}
\label{interpola}\end{center}
\end{figure}

Hence, we have solved the interpolation problem and we can use $\Lambda^*$
for fixing either $d_{0}$ or $d_{L}$, but we cannot choose both ends 
of the rulings. An example of this 
construction is shown in Figure~\ref{interpola}

The procedure for solving the problem is clear: 

\begin{enumerate}
    \item Write the algebraic equation \ref{cramer} with the
    B-spline polygon for $c(u)$, vectors $\mathbf{v}$, $\mathbf{w}$
    and the list of knots and obtain a solution $M^{*}_{0}$.  For any
    value of $\Lambda^*$ the resulting developable surface will have
    $c(u)$ as part of the boundary and the first and last rulings will
    be straight lines with respective directions $\mathbf{v}$,
    $\mathbf{w}$.

    \item  Fix $\Lambda^{*}_{0}$ by choosing either $d_{0}$ or $d_{L}$ in 
    Eq.~\ref{sigmatau}.

    \item  Use the recursivity relation in Eq.~\ref{control} for computing 
    the vertices $d_{i}$ for $d(u)$.
    
    \item The B-spline polygons $\{c_{0},\ldots, c_{L}\}$, 
$\{d_{0},\ldots,d_{L}\}$ form the B-spline net for the 
developable patch complying with the prescription.
\end{enumerate}

We illustrate this with an example, which will be useful as a first step for
following sections:

\begin{example}\label{spline3} Consider a spline curve of degree three 
and three pieces with B-spline polygon
\[c_{0}=(0, 0, 0),\ c_{1}=(2, 3, 0),\ c_{2}=(4, 3, 0),\]\[  
c_{3}=(5,0,0),\ c_{4}=(7,2,1),\ c_{5}=(9,-1,3),\]
and list of knots $\{0,0,0,0.3,0.7,1,1,1\}$, not uniformly spaced. 
For the first ruling we choose direction $\mathbf{v}=(0,0,2)$ and for 
the last ruling we choose $\mathbf{w}=(-1,0,1)$. Find a developable 
surface patch bounded by $c(u)$ and the rulings defined by $\mathbf{v}$, 
$\mathbf{w}$.\end{example}

We calculate the determinant in Eq.~\ref{cramer},
\[\det(a(M^*)-c_{L},\mathbf{v}, 
    \mathbf{w})\]\[=\frac{2({M^*}^4+6.2{M^*}^3-12.3{M^*}^2+9.3{M^*}-2.1)}{{M^*}^3(
    {M^*}-0.3)({M^*}-0.7)},\]
and we ensure developability  by choosing the parameter $M^*$ as one of 
the real solutions of
\[{M^*}^4+6.2{M^*}^3-12.3{M^*}^2+9.3{M^*}-2.1=0,\]
which are $M^{*}=-7.91,\ 0.37$.

We further choose $d_{0}=c_{0}+\mathbf{v}=(0,0,2)$ along the first ruling, 
which amounts to choosing $\sigma=1$ in Eq.~\ref{sigmatau}, to obtain 
the respective values of the parameter $\Lambda^*=-6.18, 
\ 0.61$. We perform the calculations for the first pair of parameters, 
$\Lambda^*=-6.18$, $M^{*}=-7.91$.

\begin{figure}[h]\begin{center}
\includegraphics[height=0.2\textheight]{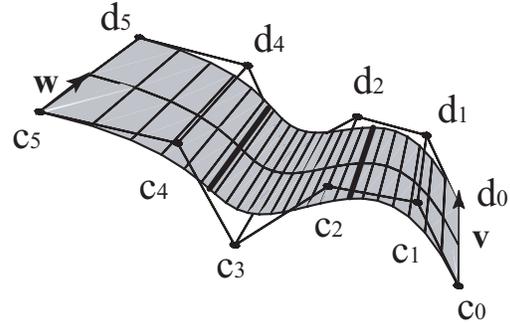}
\caption{Developable surface of degree 3 and 3 pieces}
\label{cubica}\end{center}
\end{figure}

We may use now Corollary~1 to obtain the B-spline polygon of the 
other boundary curve of the developable patch through $c(u)$ with 
prescribed rulings,
\begin{eqnarray*}
d_{i+1}\!=\!\frac{(u_{i+n}-\Lambda^*) c_{i}+(\Lambda^*\!\!\!-u_{i})  c_{i+1}+
(M^{*}\!\!\!-u_{i+n})d_{i}}{M^*-u_{i}}\end{eqnarray*} for $ i=0\ldots L-1$,
\begin{eqnarray*}
d_{1}&=&\frac{(u_{3}-\Lambda^*) c_{0}+(\Lambda^*-u_{0})  c_{1}+
(M^{*}-u_{3})d_{0}}{M^*-u_{0}}\\&=&(1.56, 2.34, 2.08)
\\
d_{2}&=&\frac{(u_{4}-\Lambda^*) c_{1}+(\Lambda^*-u_{1})  c_{2}+
(M^{*}-u_{4})d_{1}}{M^*-u_{1}}\\&=&(3.09, 2.29, 2.26)
\\
d_{3}&=&\frac{(u_{5}-\Lambda^*) c_{2}+(\Lambda^*-u_{2})  c_{3}+
(M^{*}-u_{5})d_{2}}{M^*-u_{2}}\\&=&(3.75,-0.15,2.55)
\\
d_{4}&=&\frac{(u_{6}-\Lambda^*) c_{3}+(\Lambda^*-u_{3})  c_{4}+
(M^{*}-u_{6})d_{3}}{M^*-u_{3}}\\&=&(5.22, 1.42, 3.55)
\\
d_{5}&=&\frac{(u_{7}-\Lambda^*) c_{4}+(\Lambda^*-u_{4})  c_{5}+
(M^{*}-u_{7})d_{4}}{M^*-u_{4}}\\&=&(6.76, -1.00, 5.24).
\end{eqnarray*}
and check that in fact $d_{5}$ lies on the last ruling since
\[d_{5}-c_{5}=(-2.24,0.00,2.24),\]
which is a vector proportional to $\mathbf{w}$. The resulting patch 
is shown in Figure~\ref{cubica}.

Another way to look at this developable surface would be to split the 
spline curve into three cubic B\'ezier curves, 
$\{C_{0},C_{1},C_{2},C_{3}\}$, $\{C_{3},C_{4},C_{5},C_{6}\}$, 
$\{C_{6},C_{7},C_{8},C_{9}\}$, by knot insertion, 
\[C_{0}=(0,0,0),\ C_{1}=(2,3,0),\ C_{2}=(2.86,3,0),\]
\[C_{3}=(3.48,2.61,0),\ C_{4}=(4.3,2.1,0),\]\[ C_{5}=(4.7,0.9,0),\ 
C_{6}=(5.52, 1.04, 0.33),\]
\[C_{7}=(6.14,1.14,0.57),\ C_{8}=(7,2,1),\]\[ 
C_{9}=(9,-1,3).\]

If we also split by knot insertion the other boundary curve in three
cubic pieces, 
$\{D_{0},D_{1},D_{2},D_{3}\}$, $\{D_{3},D_{4},D_{5},D_{6}\}$, 
$\{D_{6},D_{7},D_{8},D_{9}\}$, by knot insertion, 
\[D_{0}=(0,0,2),\ D_{1}=(1.56, 2.34, 2.08),\]\[ D_{2}=(2.21,2.32,2.15),\]
\[D_{3}=(2.67, 1.99, 2.24),\ D_{4}=(3.29,1.56,2.35),\]\[ D_{5}=(3.55,0.58,2.46),\ 
D_{6}=(4.15, 0.68, 2.84),\]
\[D_{6}=(4.15, 0.68, 2.84),\ D_{7}=(4.59,0.75,3.12),\]\[ D_{8}=(5.22, 1.42, 3.55),
\ D_{9}=(6.76, -1.00, 5.24),\]
it is easy to check that the three pieces of the composite ruled surface are 
in fact independent developable surfaces on their respective intervals $[0,0.3]$, 
$[0.3,0.7]$, $[0.7,1]$, with the same parameters $\Lambda^{*}=-6.18$, 
$M^{*}=0.61$. The boundary rulings of these B\'ezier developable 
surfaces have been marked in Figure~\ref{cubica}.

\section{Degree elevation of developable surfaces}
\label{elevate}
We have seen how to interpolate a spline developable surface bounded
by a spline curve and two rulings, but we cannot choose both endpoints
for such rulings.  This is a limitation of the procedure in
\cite{leonardo-developable} described in the previous sections.  A way
to deal with this problem is to try to find a solution of higher
degree.

As it is pointed out in \cite{aumann1}, degree
elevation may be used for enlarging a developable patch by modifying
the length of the ruling segments of the patch.  The idea is simple.
We may modify the length of the director vector
\[\mathbf{w}(u)=d(u)-c(u),\]
of each ruling by multiplication by a function $f(u)$,
\[\mathbf{\tilde w}(u)=f(u)\mathbf{w}(u)=\tilde d(u)-c(u),\]
and as a consequence the boundary of the surface patch changes. For 
instance the new second curve $\tilde d(u)$ starts at $\tilde 
d_{0}=c_{0}+f(u_{n-1})(d_{0}-c_{0})$ and ends at $\tilde 
d_{L}=c_{L}+f(u_{n+N-1})(d_{L}-c_{L})$.

\begin{figure}[h]\begin{center}
\includegraphics[height=0.25\textheight]{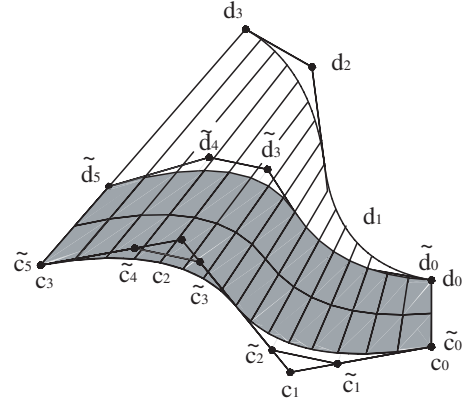}
\caption{Developable surface of degree 2 and 2 pieces stretched to 
a patch of degree 3}
\label{elevat}\end{center}
\end{figure}

It is clear that this transformation just changes the patch of the
developable surface that is covered by the parametrisation and it
allows us to change the endpoints $d_{0}$ and $d_{L}$ of the first and
last rulings.  The only problem is that the curve $\tilde d(u)$ is no
longer a spline of degree $n$.  The simplest choice for the factor is
an affine function $f(u)=au+b$, and in this case the new surface patch
\[\tilde b(u,v)=(1-u)c(u)+v\tilde d(u)\] will be of degree $(n+1,1)$. 
An example is shown in Figure~\ref{elevat}.

The next step will be the calculation of the B-spline polygon of the 
new boundary of the extended surface patch.

First, we obtain the blossom of the new parametrised curve,
\[\tilde d(u)=\left(1-f(u)\right)c(u)+f(u)d(u).\]

The blossom is a $(n+1)$-affine symmetric form $\tilde 
d[u_{0},\ldots,u_{n}]$ for which
\[\tilde d(u)=\tilde d[u^{<n+1>}].\]

Since $f(u)$ is an affine function, it is already its own
blossom, $f[u]=f(u)$. For the product $h(u)=f(u)d(u)$ it is simple to 
produce an $(n+1)$-affine form $\hat h$ satisfying $\hat h[u^{<n+1>}]=h(u)$,
\[\hat h[u_{0},\ldots,u_{n}]=f(u_{0})d[u_{1},\ldots,u_{n}],\]
but this form is clearly non-symmetric. 

However, we may obtain a symmetric form just by permuting the 
argument of the function $f$,
\[h[u_{0},\ldots,u_{n}]\!=\!\frac{\displaystyle\sum_{i=0}^n\!f(u_{i})
d[u_{0},\ldots,u_{i-1},u_{i+1},\ldots,u_{n}]}{n+1}.\]

This form $h$ is $(n+1)$-affine, symmetric and clearly 
$h[u^{<n+1>}]=h(u)$. Hence, it is the blossom of the parametrisation 
$h(u)$.

We may use this result to conclude that the blossom of $\tilde d(u)$ is 
given by
\begin{eqnarray}\!\!\!\!\!\!\!\!\!\!\!\!&&
\tilde d[u_{0},\ldots,u_{n}]\!=\!
\frac{\displaystyle\sum_{i=0}^{n}
f(u_{i})d[u_{0},\ldots,u_{i-1},u_{i+1},\ldots,u_{n}]}{n+1}
\nonumber\\\!\!\!\!\!\!\!\!\!\!\!\!&&+\frac{\displaystyle\sum_{i=0}^{n}\!
\big(1-f(u_{i})\big)c[u_{0},\ldots,u_{i-1},u_{i+1},\ldots,u_{n}]}
{n+1}.\label{elevd}
\end{eqnarray}

%

The degree of the curve $c(u)$ must be formally elevated to $n+1$ in 
order to complete the B-spline net of the surface patch of degree 
$(n+1,1)$. It can be computed by taking $f\equiv 1$ in the previous 
formula for $\tilde d$. The degree-elevated blossom for $c(u)$ is 
\[\tilde c[u_{0},\ldots,u_{n}]\!=\!\frac{1}{n+1}\!\sum_{i=0}^{n}
c[u_{0},\ldots,u_{i-1},u_{i+1},\ldots,u_{n}].\]

The list of knots of the degree-elevated curves (\cite{farin}) is also modified by 
increasing by one the multiplicity of the inner knots 
$u_{n-1}$,\ldots,$u_{n+N-1}$,
\begin{eqnarray*}\label{reknot}
\{u_{0},\ldots,u_{n-1},u_{n-1},\ldots,u_{n+N-1},u_{n+N-1},\ldots,u_{K}\}.
\end{eqnarray*}

Then the new B-spline polygons of the curves $c(u)$ and $\tilde d(u)$ 
will be $\{\tilde c_{0},\ldots,\tilde c_{L'}\}$, $\{\tilde 
d_{0},\ldots,\tilde d_{L'}\}$, 
\begin{equation}\label{revertex}
\tilde c_{i}=\tilde c[\tilde u_{i},\ldots,\tilde u_{i+n}],\ 
\tilde d_{i}=\tilde d[\tilde u_{i},\ldots,\tilde u_{i+n}],\end{equation}
for $i=0,\ldots,L'$. The list of knots has been renumbered as $\{\tilde 
u_{0},\ldots,\tilde u_{K'}\}$ in order to have correlative indices.

This construction is useful to solve the following interpolation 
problem:

\noindent \textbf{Problem 2:} Given a spline curve $c(u)$ of degree $n$, $N$ 
    pieces,     B-spline polygon $\{c_{0},\ldots,c_{L}\}$ and list of knots 
    $\{u_{0},\ldots,u_{K}\}$, $u\in[a,b]$, $a=u_{n-1}$, 
    $b=u_{n+N-1}$, and 
two points $d_{0}$, $d_{L}$, find
a developable surface $b(u,v)$ such that $c(u,0)=c(u)$,
$c(a,1)=d_{0}$, $c(b,1)=d_{L}$.

The procedure for solving this problem is clear: 

\begin{enumerate}
    \item  Write the algebraic equation \ref{cramer} with the 
    B-spline polygon for $c(u)$, the list of knots and vectors for 
    the rulings $\overrightarrow{c_{0}d_{0}}$, 
    $\overrightarrow{c_{L}d_{L}}$ and obtain a 
    solution $M^{*}_{0}$.

    \item  Fix $\Lambda^{*}_{0}$ by choosing $d_{0}$ in 
    Eq.~\ref{sigmatau} ($\sigma=1$, but $\tau\neq 1$ in general).

    \item  Use the recursivity relation in Eq.~\ref{control} for computing 
    the vertices of $d(u)$.
    
    \item Increase by one the multiplicity of the inner knots of the 
    boundary curves.
    
    \item Formally raise the degree of $c(u)$ and compute the new 
    B-spline vertices $\tilde c_{i}$ with Eq.~\ref{elevboor}.
    
    \item Choose 
$f(u)$ so that $f(a)=1$, $f(b)=1/\tau$,
\begin{eqnarray}f(u)=\frac{b-u}{b-a}+\frac{1}{\tau}\frac{u-a}{b-a}.
\label{fu}\end{eqnarray}

\item Use this function to compute the B-spline vertices $\tilde 
d_{i}$ for the new boundary curve $\tilde d(u)$ with 
Eq.~\ref{revertex} and Eq.~\ref{elevd}.

\item The B-spline polygons $\{\tilde c_{0},\ldots,\tilde c_{L'}\}$, 
$\{\tilde d_{0},\ldots,\tilde d_{L'}\}$ form the B-spline net for the 
developable patch complying with the prescription.

\end{enumerate}


We go back now to Example \ref{spline3}: 

\begin{example}\label{spline4} Consider a spline curve of degree three
and three pieces with B-spline polygon \[c_{0}=(0, 0, 0),\ c_{1}=(2,
3, 0),\ c_{2}=(4, 3, 0),\]\[ c_{3}=(5,0,0),\ c_{4}=(7,2,1),\
c_{5}=(9,-1,3),\] and list of knots $\{0,0,0,0.3,0.7,1,1,1\}$.
For the first ruling we choose direction
$\mathbf{v}=(0,0,2)$ and for the last ruling we choose
$\mathbf{w}=(-1,0,1)$.  Find a developable surface patch bounded by
$c(u)$, an unknown curve $\tilde d(u)$ and the rulings defined by
$\mathbf{v}$, $\mathbf{w}$, such that $\tilde d(0)=c_{0}+\mathbf{v}=(0,0,2)$,
$\tilde d(1)=c_{5}+\mathbf{w}=(8,-1,4)$.\end{example}

We already have obtained that the spline curve with B-spline polygon 
\[d_{0}=(0, 0, 2),\ d_{1}=(1.56, 2.34, 2.08),\]\[ 
d_{2}=(3.09, 2.29, 2.26),\ d_{3}=(3.75,-0.15,
2.55),\]\[  d_{4}=(5.22, 1.42, 3.55),\ 
d_{5}=(6.76, -1.00, 5.24),\]
and the same list of knots provides a developable surface patch with 
the required prescription except that $d_{5}$ lies on the final 
ruling, but it is not $(8,-1,4)$. In fact, $d_{5}=c_{5}+\tau 
\mathbf{w}$ with $\tau=2.24$.

In order to shorten the surface patch so that the final vertex 
of the new boundary curve $\tilde d(u)$ is $(8,-1,4)$, we have to 
raise the degree of the curves from three to four.

Increasing the multiplicity of the inner knots 0, 0.3, 0.7, 1, we get 
the new list of knots for the degree-elevated curves,
\[\{0,0,0,0,0.3,0.3,0.7,0.7,1,1,1,1\}.\]

We calculate first the B-spline polygon for $c(u)$ as a curve of 
formal degree four  with Eq.~\ref{elevboor}. The auxiliary points are 
computed in Appendix \ref{append}.
\begin{eqnarray*}
    \tilde c_{0}&=&\tilde c[0,0,0,0]=c[0,0,0]=(0,0,0)\\
\tilde c_{1}&=&\tilde 
c[0,0,0,0.3]=\frac{c[0,0,0]+3c[0,0,0.3]}{4}\\&=&(1.5,2.25,0)\\
\tilde c_{2}&=&\tilde 
c[0,0,0.3,0.3]=\frac{c[0,0,0.3]+c[0,0.3,0.3]}{2}
\\&=&(2.43, 3, 0)\\
\tilde c_{3}&=&\tilde 
c[0,0.3,0.3,0.7]\\&=&\frac{c[0,0.3,0.3]+2c[0,0.3,0.7]+c[0.3,0.3,0.7]}{4}
\\&=&(3.79, 2.78, 0)\end{eqnarray*}
\begin{eqnarray*}
\tilde c_{4}&=&\tilde 
c[0.3,0.3,0.7,0.7]\\&=&\frac{c[0.3,0.3,0.7]+c[0.3,0.7,0.7]}{2}
=(4.5, 1.5, 0)\\
\tilde c_{5}&=&\tilde 
c[0.3,0.7,0.7,1]\\&=&\frac{c[0.3,0.7,0.7]+2c[0.3,0.7,1]+c[0.7,0.7,1]}{4}
\\&=&(5.21, 0.51, 0.14)
\\
\tilde c_{6}&=&\tilde c[0.7,0.7,1,1]=\frac{c[0.7,0.7,1]+c[0.7,1,1]}{2}
\\&=&
(6.57, 1.57, 0.79)\\
\tilde c_{7}&=&\tilde 
c[0.7,1,1,1]=\frac{3c[0.7,1,1]+c[1,1,1]}{4}
\\&=&(7.5,1.25,1.5)\\
\tilde c_{8}&=&\tilde c[1,1,1,1]=c[1,1,1]=(9,-1,3).
\end{eqnarray*}

Now we have to move the curve $d(u)$ over the developable surface
patch so that the new boundary curve $\tilde d(u)$ goes through the
endpoints of both rulings, shortening the director vector 
$\mathbf{w}(u)$ by a factor $f(u)$ as in Eq.~\ref{fu},
\[f(u)=(1-u)+\frac{u}{2.24}.\]

Finally, we use Eq.~\ref{elevd} to compute the B-spline polygon of the 
new boundary curve of degree four that goes through the endpoints of 
both rulings,
\begin{eqnarray*}
\tilde d_{0}\!\!\!\!&=&\!\!\!\!\tilde d[0,0,0,0]=f(0)d[0,0,0]+(1-f(0))c[0,0,0]
\\\!\!\!\!&=&\!\!\!\!d_{0}=(0, 0, 2)
\\
\tilde d_{1}\!\!\!\!&=&\!\!\!\!\tilde d[0,0,0,0.3]=
\frac{f(0.3)d[0,0,0]+3f(0)d[0,0,0.3]}{4}\\\!\!\!\!&+&\!\!\!\!\frac{(1-f(0.3))c[0,0,0]
3(1-f(0)c[0,0,0.3]}{4}
\\\!\!\!\!&=&\!\!\!\!(1.17, 1.76, 1.97)
\\
\tilde d_{2}\!\!\!\!&=&\!\!\!\!\tilde d[0,0,0.3,0.3]=
\frac{f(0.3)d[0,0,0.3]+f(0)d[0,0.3,0.3]}{2}\\\!\!\!\!&+&\!\!\!\!\frac{(1-f(0.3))c[0,0,0.3]+
(1-f(0)c[0,0.3,0.3]}{2}
\\\!\!\!\!&=&\!\!\!\!(1.93, 2.39, 1.94)
\\
\tilde d_{3}\!\!\!\!&=&\!\!\!\!\tilde d[0,0.3,0.3,0.7]=
\frac{f(0.7)d[0,0.3,0.3]}{4}\\\!\!\!\!\!&+&\!\!\!\!\!\frac{2f(0.3)d[0,0.3,0.7]+
f(0)d[0.3,0.3,0.7]}{4}\\\!\!\!\!\!&+&\!\!\!\!\!\frac{(1-f(0.7))c[0,0.3,0.3]}{4}\\
\!\!\!\!&+&\!\!\!\!\frac{(1-f(0.3))c[0,0.3,0.7]}{2}
\\\!\!\!\!\!&+&\!\!\!\!\!\frac{(1-f(0))c[0.3,0.3,0.7]}{4}
=(3.06, 2.24, 1.86)
\end{eqnarray*}
\begin{eqnarray*}
\tilde d_{4}\!\!\!\!&=&\!\!\!\!\tilde d[0.3,0.3,0.7,0.7]=
\frac{f(0.3)d[0.3,0.7,0.7]}{2}\\\!\!\!\!\!&+&\!\!\!\!\!\frac{f(0.7)d[0.3,0.3,0.7]
+(1-f(0.3))c[0.3,0.7,0.7]}{2}\\\!\!\!\!\!&+&\!\!\!\!\!\frac{(1-f(0.7))c[0.3,0.3,0.7]}{2}
=(3.71, 1.20, 1.74)
\\
\tilde d_{5}\!\!\!\!&=&\!\!\!\!\tilde d[0.3,0.7,0.7,1]=
\frac{f(1)d[0.3,0.7,0.7]}{4}\\\!\!\!\!\!&+&\!\!\!\!\!\frac{2f(0.7)d[0.3,0.7,1]
+f(0.3)d[0.7,0.7,1]}{4}\\\!\!\!\!\!&+&\!\!\!\!\!\frac{
(1-f(1))c[0.3,0.7,0.7]}{4}\\\!\!\!\!\!&+&\!\!\!\!\!\frac{(1-f(0.7))c[0.3,0.7,1]
}{2}\\\!\!\!\!\!&+&\!\!\!\!\!\frac{(1-f(0.3))c[0.7,0.7,1]}{4}
=(4.38, 0.35, 1.73)
\\
\tilde d_{6}\!\!\!\!&=&\!\!\!\!\tilde d[0.7,0.7,1,1]=
\frac{f(1)d[0.7,0.7,1]}{2}\\\!\!\!\!\!&+&\!\!\!\!\!\frac{f(0.7)d[0.7,1,1]+
(1-f(1))c[0.7,0.7,1]}{2}\\\!\!\!\!\!&+&\!\!\!\!\!\frac{(1-f(0.7))c[0.7,1,1]}{2}
=(5.68, 1.30, 2.14)
\\
\tilde d_{7}\!\!\!\!&=&\!\!\!\!\tilde d[0.7,1,1,1]=
\frac{3f(1)d[0.7,1,1]+f(0.7)d[1,1,1]}{4}\\\!\!\!\!&+&\!\!\!\!\frac{3(1-f(1))c[0.7,1,1]+(1-f(0.7))c[1,1,1]}{4}
\\\!\!\!\!&=&\!\!\!\!(6.56, 1.05, 2.70)
\\
\tilde d_{8}\!\!\!\!&=&\!\!\!\!\tilde d[1,1,1,1]=f(1)d[1,1,1]\\
\!\!\!\!&+&\!\!\!\!(1-f(1))c[1,1,1]=(8, -1, 4).
\end{eqnarray*}

\begin{figure}[h]\begin{center}
\includegraphics[height=0.2\textheight]{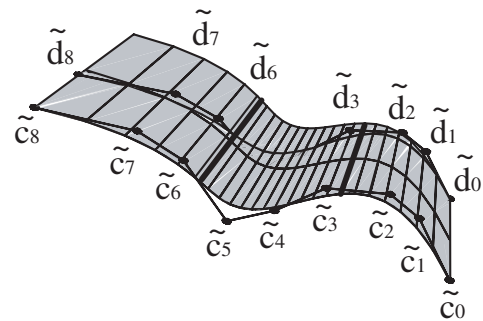}
\caption{Degree-elevation and restriction of the developable surface 
patch in Figure~\ref{cubica}}
\label{cuartica}\end{center}
\end{figure}

The degree-elevated B-spline net for the new surface patch, complying with the requirements of 
the example can be seen in Figure~\ref{cuartica}.

We could also have split the original curve $c(u)$ in three cubic B\'ezier pieces and raise 
the degree of each of them to obtain curves of formally degree four 
with control points,
\[\tilde C_{0}=(0,0,0),\ \tilde C_{1}=(1.5,2.25,0),\ \tilde 
C_{2}=(2.43,3,0),\]\[ \tilde C_{3}=(3.01, 2.90, 0),\ 
\tilde C_{4}=(3.48,2.61,0),\]\[ \tilde C_{5}=(4.09,2.23, 0),\ \tilde 
C_{6}=(4.5, 1.5, 0),\]\[ \tilde C_{7}=(4.91 0.93, 0.08),\ 
\tilde C_{8}=(5.52, 1.04, 0.33),\]
\[\tilde C_{9}=(5.99,1.12, 0.51),\ \tilde 
C_{10}=(6.57, 1.57, 0.79),\]\[ \tilde C_{11}=(7.5, 1.25, 1.5),\ 
\tilde C_{12}=(9,-1,3),\]
and use the construction in \cite{aumann1} to extend each B\'ezier developable 
surface patch to comply with the prescription of endpoints, 
by multiplication by the same factor $f(u)$. One reaches the 
same result as applying insertion of knots $0,0.3,0.7,1$ to $\tilde d(u)$,
\[\tilde D_{0}=(0,0,2),\ \tilde D_{1}=(1.17, 1.76, 1.97),\]\[ \tilde 
D_{2}=(1.93, 2.39, 1.94),\ \tilde D_{3}=(2.41, 2.32, 1.91),\]\[ 
\tilde D_{4}=(2.81, 2.10, 1.87),\ \tilde D_{5}=(3.34, 1.79, 1.81),\]\[ \tilde 
D_{6}=(3.71, 1.20, 1.74),\ \tilde D_{7}=(4.09, 0.71, 1.74),\]\[ 
\tilde D_{8}=(4.68, 0.82, 1.86),\ \tilde D_{9}=(5.12, 0.89, 1.96),\]\[ \tilde 
D_{10}=(5.68, 1.30, 2.14),\ \tilde D_{11}=(6.56, 1.05, 2.70),\]\[ 
\tilde D_{12}=(8,-1,4).\]

The boundary rulings of the quartic B\'ezier developable 
surfaces have been marked in Figure~\ref{cuartica}.

\section{Triangular developable surfaces}
\label{triangles}

We may pose another interpolation problem in which the first ruling 
collapses to a point, $c(a)=d(a)$,
\[b(u,v)=(1-v)c(u)+vd(u),\quad u\in [a,b].\] The resulting developable patch is triangular 
in the sense that it is bounded by two curves and just one straight 
segment. Instead of the first point of the unknown curve of the 
boundary, we may give as datum its initial velocity $d'(a)$.

\noindent \textbf{Problem 3:} Given a spline curve $c(u)$ of degree $n$, $N$ 
    pieces, 
    B-spline polygon $\{c_{0},\ldots,c_{L}\}$ and list of knots 
    $\{u_{0},\ldots,u_{K}\}$, $u\in[a,b]$, $a=u_{n-1}$, 
    $b=u_{n+N-1}$, a point $d_{L}$ and a vector $d'(a)$, 
    find a triangular developable surface
$b(u,v)$ through $c(u)$, such that $c(u,0)=c(u)$, $c(a,v)=c_{0}$ 
for all $v$, $c(b,1)=d_{L}$, $c_{u}(a,1)=d'(a)$.

We do not know the first ruling of the surface, but we may use
previous constructions to compute a spline developable patch through 
the curve $c(u)$ and use $d_{L}$ to fix the last ruling, 
\[b(u,v)=c(u)+v \mathbf{w}(u),\quad \mathbf{w}(u)=d(u)-c(u).\]

In order to collapse the first ruling to a point, we shorten 
the patch along the rulings,
\begin{equation}\label{shorten}\hat b(u,v)=c(u)+vf(u)\mathbf{w}(u),\quad f(u)=
\frac{u-a}{b-a},\end{equation}
so that $\hat c(a,v)=c_{0}$ for all $v$.

We compute the velocity,
\[\hat 
b_{u}(u,v)=c'(u)+\frac{v}{b-a}\mathbf{w}(u)+vf(u)\mathbf{w}'(u),\]
of the boundary curve $d(u)$ at $u=a$, making use of Eq.~\ref{velo}
\begin{eqnarray*}\hat d'(a)&=&\hat c_{u}(a,1)=c'(a)+ 
\frac{\mathbf{w}(a)}{b-a}\\&=&n\frac{c_{1}-c_{0}}{u_{n}-u_{n-1}}
+\frac{d_{0}-c_{0}}{b-a} \end{eqnarray*}
and from this expression we get the vertex $d_{0}$ that is necessary 
for obtaining the velocity $\hat d'(a)$,
\begin{eqnarray}\label{veloc}
d_{0}=c_{0}+(b-a)\left(\hat d'(a)-
n\frac{c_{1}-c_{0}}{u_{n}-u_{n-1}}\right),\end{eqnarray}

Since we need to fix both $d_{0}$ and $d_L$ to obtain the developable 
patch $b(u,v)$, the construction from the previous section is 
required and hence such a patch must be of degree $n+1$. Since $c(u)$ 
is still of degree $n$, the calculation done in Eq.~\ref{veloc} is 
nonetheless valid whereas we keep the original vertices $c_{0}$ and 
$c_{1}$. Finally, shortening the surface patch as in Eq.~\ref{shorten} with $f(u)$ 
produces a triangular patch of degree $n+2$.

Summarising, the solution of this problem is reduced to the one of Problem~2: 

\begin{enumerate}
    
\item Calculate the vertex $d_{0}$ and $\mathbf{v}=d_{0}-c_{0}$ using Eq.~\ref{veloc}.

\item  Write the algebraic equation \ref{cramer} with the 
    B-spline polygon for $c(u)$, the list of knots and vectors for 
    the rulings $\overrightarrow{c_{0}d_{0}}$, 
    $\overrightarrow{c_{L}d_{L}}$ and obtain a 
    solution $M^{*}_{0}$.

    \item  Fix $\Lambda^{*}_{0}$ by choosing $d_{0}$ in 
    Eq.~\ref{sigmatau} ($\sigma=1$, but $\tau\neq 1$ in general).

    \item  Use the recursivity relation in Eq.~\ref{control} for computing 
    the vertices of $d(u)$.
    
    \item Increase by one the multiplicity of the inner knots of the 
    boundary curves.

    \item Formally raise the degree of $c(u)$ and compute the new 
    B-spline vertices $\tilde c_{i}$ with Eq.~\ref{elevboor}.
    
    \item Choose 
$f(u)$ so that $f(a)=1$, $f(b)=1/\tau$,
\begin{eqnarray*}f(u)=\frac{b-u}{b-a}+\frac{1}{\tau}\frac{u-a}{b-a}.
\end{eqnarray*}

\item Use this function to compute the B-spline vertices $\tilde 
d_{i}$ for the new boundary curve $\tilde d(u)$ with 
Eq.~\ref{revertex} and Eq.~\ref{elevd}.

    \item Increase by one the multiplicity of the inner knots of the 
    boundary curves.

\item  Formally raise the degree of $\tilde c(u)$ and compute the new 
B-spline vertices $\hat c_{i}$ with Eq.~\ref{elevboor}.

\item Use a function $\hat f(u)=u$ to shrink the first ruling to a 
point and compute the B-spline vertices $\hat 
d_{i}$ for the new boundary curve $\hat d(u)$ with 
Eq.~\ref{revertex} and Eq.~\ref{elevd}.

\item The B-spline polygons $\{\hat c_{0},\ldots,\hat c_{L'}\}$, 
$\{\hat d_{0},\ldots,\hat d_{L'}\}$ form the B-spline net for the 
triangular developable patch complying with the prescription.

\end{enumerate}



\begin{example}\label{splinet} Consider a spline curve of degree three
and three pieces with B-spline polygon \[c_{0}=(0, 0, 0),\ c_{1}=(2,
3, 0),\ c_{2}=(4, 3, 0),\]\[ c_{3}=(5,0,0),\ c_{4}=(7,2,1),\
c_{5}=(9,-1,3),\] and list of knots $\{0,0,0,0.3,0.7,1,1,1\}$.  
For the last ruling we choose direction
$\mathbf{w}=(-1,0,1)$.  Find a triangular developable surface patch
bounded by $c(u)$, an unknown curve $\hat d(u)$ and the ruling
defined by $\mathbf{w}$, such that $\hat d(0)=c_{0}$, 
$\hat d'(0)= (20,30.5, 2)$, $\hat d(1)=c_{5}+\mathbf{w}=(8,-1,4)$.\end{example}

First of all, we calculate the first ruling of the developable
surface.  According to Eq.~\ref{veloc} we need
\[\mathbf{v}=d_{0}-c_{0}=\hat
d'(0)+\frac{3}{0.3}(c_{0}-c_{1})=(0,0.5,2),\] and we
calculate the determinant in Eq.~\ref{cramer},
\[\det(a(M^*)-c_{L},\mathbf{v},
\mathbf{w})\]\[=\frac{8{M^*}^4+2.6{M^*}^3-16{M^*}^2+14.5{M^*}-3.5}{{M^*}^3(
{M^*}-0.3)({M^*}-0.7)},\] 
so that developability is granted by
choosing parameter $M^*$ as a real solution of
\[8{M^*}^4+2.6{M^*}^3-16{M^*}^2+14.5{M^*}-3.5=0,\] that is 
$M^{*}=-1.92,\ 0.38$. The other two solutions are complex.

For having $d_{0}=(0,0.5,2)$ on the first
ruling, we need to take $\sigma=1$ in Eq.~\ref{sigmatau}.  The
respective values of parameter $\Lambda^*$ are 
$-1.16$, $0.59$. 
We choose the first pair of parameters for our calculations,
$\Lambda_{0}^*=-1.16$, $M_{0}^{*}=0.59$.
\begin{figure}[h]\begin{center}
\includegraphics[height=0.2\textheight]{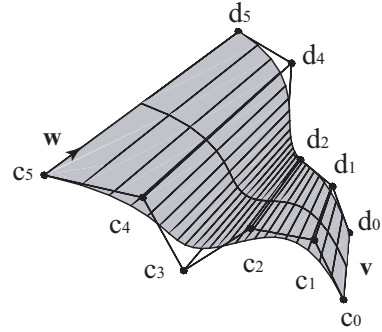}
\caption{Developable surface of degree 3 and 3 pieces}
\label{tcubica}\end{center}
\end{figure}
We calculate next the B-spline polygon for the second boundary curve 
according to Corollary~1,
\begin{eqnarray*}
d_{i+1}\!=\!\frac{(u_{i+n}-\Lambda^*) c_{i}\!+\!(\Lambda^*-u_{i})  
c_{i+1}\!+
\!(M^{*}-u_{i+n})d_{i}}{M^*-u_{i}}\end{eqnarray*} for $i=0\ldots L-1$.
\begin{eqnarray*}
d_{0}&=&(0, 0.5, 2)\\
d_{1}&=&\frac{(u_{3}-\Lambda^*) c_{0}+(\Lambda^*-u_{0})  c_{1}+
(M^{*}-u_{3})d_{0}}{M^*-u_{0}}\\&=&(1.21, 2.39, 2.31)
\\
d_{2}&=&\frac{(u_{4}-\Lambda^*) c_{1}+(\Lambda^*-u_{1})  c_{2}+
(M^{*}-u_{4})d_{1}}{M^*-u_{1}}\\&=&(2.13, 2.17, 3.16)
\\
d_{3}&=&\frac{(u_{5}-\Lambda^*) c_{2}+(\Lambda^*-u_{2})  c_{3}+
(M^{*}-u_{5})d_{2}}{M^*-u_{2}}\\&=&(1.77, -0.07, 4.80)
\\
d_{4}&=&\frac{(u_{6}-\Lambda^*) c_{3}+(\Lambda^*-u_{3})  c_{4}+
(M^{*}-u_{6})d_{3}}{M^*-u_{3}}\\&=&(2.07, 1.22, 6.97)
\\
d_{5}&=&\frac{(u_{7}-\Lambda^*) c_{4}+(\Lambda^*-u_{4})  c_{5}+
(M^{*}-u_{7})d_{4}}{M^*-u_{4}}\\&=&(2.92, -1.00, 9.08).
\end{eqnarray*}

Hence, $d_{5}-c_{5}=\tau\mathbf{w}$, with $\tau=6.08$.  We
show the surface patch in Figure~\ref{tcubica}.

\begin{figure}[h]\begin{center}
\includegraphics[height=0.2\textheight]{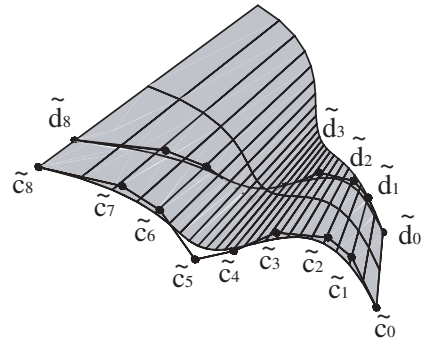}
\caption{Restriction of the developable surface 
patch in Figure~\ref{tcubica}}
\label{tcuartica}\end{center}
\end{figure}

Next we shorten the surface patch so that the new boundary curve 
$\hat d(u)$ ends up at $(8,-1,4)$. From the previous example we know 
that we are to increase the multiplicity of the inner knots by one,
\[\{0,0,0,0,0.3,0.3,0.7,0.7,1,1,1,1\},\]
and formally raise the degree of $c(u)$ to four,
\[\tilde c_{0}=(0, 0, 0),\ \tilde c_{1}=(1.5,2.25,0),\ \tilde 
 c_{2}=(2.43, 3, 0),\]\[ \tilde c_{3}=(3.79, 2.78, 0),
  \ \tilde c_{4}=(4.5, 1.5, 0),\]\[
\tilde c_{5}=(5.21, 0.51, 0.14),\ 
\tilde c_{6}=
(6.57, 1.57, 0.79),\]\[ 
\tilde c_{7}
=(7.5,1.25,1.5), \ 
\tilde c_{8}=(9,-1,3),\]
and shorten the director vector 
$\mathbf{w}(u)$ by a factor $f(u)$ as in Eq.~\ref{fu},
\[f(u)=(1-u)+\frac{u}{6.08},\]
so that the new boundary curve $\tilde d(u)$ has degree four and B-spline 
polygon using Eq.~\ref{elevd}, given by
\begin{eqnarray*}
\tilde d_{0}\!\!\!\!\!&=&\!\!\!\!\!\tilde d[0,0,0,0]=f(0)d[0,0,0]+(1-f(0))
c[0,0,0]
\\ \!\!\!\!\!&=&\!\!\!\!\!d_{0}=(0, 0.5, 2)
\\
\tilde d_{1}\!\!\!\!\!&=&\!\!\!\!\!\tilde d[0,0,0,0.3]=
\frac{f(0.3)d[0,0,0]+3f(0)d[0,0,0.3]}{4}\\
\!\!\!\!\!&+&\!\!\!\!\!\frac{(1-f(0.3))c[0,0,0]+
3(1-f(0)c[0,0,0.3]}{4}
\\ \!\!\!\!\!&=&\!\!\!\!\!(0.91, 1.89, 2.11)
\\
\tilde d_{2}\!\!\!\!\!&=&\!\!\!\!\!\tilde d[0,0,0.3,0.3]=
\frac{f(0.3)d[0,0,0.3]+f(0)d[0,0.3,0.3]}{2}\\\!\!\!\!\!&+&\!\!\!\!\!
\frac{(1-f(0.3))c[0,0,0.3]+
(1-f(0)c[0,0.3,0.3]}{2}
\\\!\!\!\!\!&=&\!\!\!\!\!(1.51, 2.42, 2.20)
\\
\tilde d_{3}\!\!\!\!\!&=&\!\!\!\!\!\tilde d[0,0.3,0.3,0.7]=
\frac{f(0.7)d[0,0.3,0.3]}{4}\\\!\!\!\!\!&+&\!\!\!\!\!\frac{2f(0.3)d[0,0.3,0.7]+f(0)d[0.3,0.3,0.7]}{4}\\
\!\!\!\!\!&+&\!\!\!\!\!\frac{
(1-f(0.7))c[0,0.3,0.3]+2(1-f(0.3))c[0,0.3,0.7]}{4}
\\\!\!\!\!\!&+&\!\!\!\!\!\frac{(1-f(0))c[0.3,0.3,0.7]}{4}
=(2.39, 2.24, 2.37)
\end{eqnarray*}
\begin{eqnarray*}
\tilde d_{4}\!\!\!\!\!&=&\!\!\!\!\!\tilde d[0.3,0.3,0.7,0.7]=
\frac{f(0.3)d[0.3,0.7,0.7]}{2}\\\!\!\!\!\!&+&\!\!\!\!\!
\frac{f(0.7)d[0.3,0.3,0.7]+
(1-f(0.3))c[0.3,0.7,0.7]}{2}\\\!\!\!\!\!&=&\!\!\!\!\!\frac{(1-f(0.7))c[0.3,0.3,0.7]}{2}
=(2.97, 1.26, 2.37)
\\
\tilde d_{5}\!\!\!\!\!&=&\!\!\!\!\!\tilde d[0.3,0.7,0.7,1]=
\frac{f(1)d[0.3,0.7,0.7]}{4}\\
\!\!\!\!\!&+&\!\!\!\!\!\frac{2f(0.7)d[0.3,0.7,1]+f(0.3)d[0.7,0.7,1]}{4}\\
\!\!\!\!\!&+&\!\!\!\!\!\frac{
(1-f(1))c[0.3,0.7,0.7]}{4}\\\!\!\!\!\!&+&\!\!\!\!\!\frac{2(1-f(0.7))c[0.3,0.7,1]+(1-f(0.3))c[0.7,0.7,1]}{4}
\\\!\!\!\!\!&=&\!\!\!\!\!(3.64, 0.39, 2.34)
\\
\tilde d_{6}\!\!\!\!\!&=&\!\!\!\!\!\tilde d[0.7,0.7,1,1]=
\frac{f(1)d[0.7,0.7,1]+f(0.7)d[0.7,1,1]}{2}\\\!\!\!\!\!&+&\!\!\!\!\!\frac{
(1-f(1))c[0.7,0.7,1]+(1-f(0.7))c[0.7,1,1]}{2}
\\\!\!\!\!\!&=&\!\!\!\!\!(5.20, 1.37, 2.48)
\\
\tilde d_{7}\!\!\!\!\!&=&\!\!\!\!\!\tilde d[0.7,1,1,1]=
\frac{3f(1)d[0.7,1,1]+f(0.7)d[1,1,1]}{4}\\
\!\!\!\!\!&+&\!\!\!\!\!\frac{3(1-f(1))c[0.7,1,1]+(1-f(0.7))c[1,1,1]}{4}
\\\!\!\!\!\!&=&\!\!\!\!\!(6.26, 1.15, 2.87)
\\
\tilde d_{8}\!\!\!\!\!&=&\!\!\!\!\!\tilde d[1,1,1,1]=
f(1)d[1,1,1]+(1-f(1))c[1,1,1]\\
\!\!\!\!\!&=&\!\!\!\!\!(8, -1, 4),
\end{eqnarray*}
where the auxiliary points are computed with blossoms in 
Appendix~\ref{tappend}. The result of this restriction of the surface 
patch is shown in Figure~\ref{tcuartica}.

Finally, following Eq.~\ref{shorten}, we further trim the surface patch bounded by
$c(u)$ and $\tilde d(u)$ to shrink the first ruling to the vertex 
$c_{0}$.

Since we are raising the degree of the curves from four to five, we 
have to increase the multiplicity of the inner knots by one,
\[\{0,0,0,0,0,0.3,0.3,0.3,0.7,0.7,0.7,1,1,1,1,1\}.\]

The curve $c(u)$ becomes formally of degree five using
Eq.~\ref{elevboor} with B-spline polygon, \begin{eqnarray*}\hat
c_{0}\!\!\!\!\!&=&\!\!\!\!\!\hat c[0,0,0,0,0]=\tilde c[0,0,0,0]=(0,0,0)\\
\hat c_{1}\!\!\!\!\!&=&\!\!\!\!\!\hat 
c[0,0,0,0,0.3]=\frac{\tilde c[0,0,0,0]+4\tilde c[0,0,0,0.3]}{5}
\\ \!\!\!\!\!&=&\!\!\!\!\!
(1.20, 1.80, 0.0)\\
\hat c_{2}\!\!\!\!\!&=&\!\!\!\!\!\hat 
c[0,0,0,0.3,0.3]=\frac{2\tilde c[0,0,0,0.3]}{5}\\\!\!\!\!\!&+&\!\!\!\!\!
\frac{3\tilde c[0,0,0.3,0.3]}{5}
=(2.06, 2.70, 0.0)\\
\hat c_{3}\!\!\!\!\!&=&\!\!\!\!\!\hat 
c[0,0,0.3,0.3,0.3]=\frac{3\tilde c[0,0,0.3,0.3]}{5}\\\!\!\!\!\!&+&\!\!\!\!\!
\frac{2\tilde c[0,0.3,0.3,0.3]}{5}
=(2.66, 2.96, 0.0)\\
\hat c_{4}\!\!\!\!\!&=&\!\!\!\!\!\hat 
c[0,0.3,0.3,0.3,0.7]=\frac{\tilde c[0,0.3,0.3,0.3]
}{5}\\\!\!\!\!\!&+&\!\!\!\!\!
\frac{3\tilde c[0,0.3,0.3,0.7]+
\tilde c[0.3,0.3,0.3,0.7]}{5}
\\\!\!\!\!\!&=&\!\!\!\!\!
(3.69, 2.69, 0.0)\\
\hat c_{5}\!\!\!\!\!&=&\!\!\!\!\!\hat 
c[0.3,0.3,0.3,0.7,0.7]=\frac{2\tilde c[0.3,0.3,0.3,0.7]}{5}\\\!\!\!\!\!&+&\!\!\!\!\!
\frac{3\tilde 
c[0.3,0.3, 0.7,0.7]}{5}
=(4.34, 1.79, 0.0)\\
\hat c_{6}\!\!\!\!\!&=&\!\!\!\!\!\hat 
c[0.3,0.3,0.7,0.7,0.7]=\frac{3\tilde c[0.3,0.3,0.7,0.7]}{5}\\\!\!\!\!\!&+&\!\!\!\!\!
\frac{2\tilde 
c[0.3,0.7, 0.7,0.7]}{5}
=(4.66, 1.27, 0.03)\\
\hat c_{7}\!\!\!\!\!&=&\!\!\!\!\!\hat 
c[0.3,0.7,0.7,0.7,1]=\frac{\tilde c[0.3,0.7,0.7,0.7]}{5}\\\!\!\!\!\!&+&\!\!\!\!\!
\frac{3\tilde c[0.3,0.7,0.7,1]+
\tilde c[0.7,0.7,0.7,1]}{5}
\\\!\!\!\!\!&=&\!\!\!\!\!
(5.31, 0.72, 0.20)
\\
\hat c_{8}\!\!\!\!\!&=&\!\!\!\!\!\hat c[0.7,0.7,0.7,1,1]=\frac{2\tilde c[0.7,0.7,0.7,1]
}{5}\\\!\!\!\!\!&+&\!\!\!\!\!
\frac{3\tilde c[0.7,0.7,1,1]}{5}=
(6.34, 1.39, 0.68)\\
\hat c_{9}\!\!\!\!\!&=&\!\!\!\!\!\hat 
c[0.7,0.7,1,1,1]=\frac{3\tilde c[0.7,0.7,1,1]}{5}\\\!\!\!\!\!&+&\!\!\!\!\!
\frac{2\tilde c[0.7,1,1,1]}{5}
=(6.94, 1.44, 1.07)\\
\hat c_{10}\!\!\!\!\!&=&\!\!\!\!\!\hat 
c[0.7,1,1,1,1]=\frac{4\tilde c[0.7,1,1,1]+\tilde c[1,1,1,1]}{5}
\\\!\!\!\!\!&=&\!\!\!\!\!
(7.80, 0.80, 1.80)\\
\hat c_{11}\!\!\!\!\!&=&\!\!\!\!\!\hat c[1,1,1,1,1]=\tilde c[1,1,1,1]=(9,-1,3),
\end{eqnarray*}
and following Eq.~\ref{shorten}, we shrink the rulings  with a factor $\hat 
f(u)=u$. The auxiliary points are computed using the multiaffinity 
property of blossoms in Appendix~\ref{ttappend}.

Making use of Eq.~\ref{elevd}, we obtain the B-spline polygon of the 
final boundary curve $\hat d(u)$ of degree five,
\begin{eqnarray*}
\hat d_{0}\!\!\!\!\!&=&\!\!\!\!\!\hat d[0,0,0,0,0]=\hat f(0)\tilde d[0,0,0,0]
\\\!\!\!\!\!&+&\!\!\!\!\!
(1-\hat 
f(0))\tilde c[0,0,0,0]=\tilde c_{0}=(0, 0, 0)
\\
\hat d_{1}\!\!\!\!\!&=&\!\!\!\!\!\hat d[0,0,0,0,0.3]=
\frac{\hat f(0.3)\tilde d[0,0,0,0]}{5}\\\!\!\!\!\!&+&\!\!\!\!\!
\frac{4\hat f(0)\tilde d[0,0,0,0.3]+
(1-\hat f(0.3))\tilde c[0,0,0,0]}{5}\\\!\!\!\!\!&+&\!\!\!\!\!
\frac{4(1-\hat f(0)\tilde c[0,0,0,0.3]}{5}
\\\!\!\!\!\!&=&\!\!\!\!\!(1.20, 1.83, 0.12)
\\
\hat d_{2}\!\!\!\!\!&=&\!\!\!\!\!\hat d[0,0,0,0.3,0.3]=
\frac{2\hat f(0.3)\tilde d[0,0,0,0.3]}{5}\\\!\!\!\!\!&+&\!\!\!\!\!
\frac{3\hat f(0)\tilde 
d[0,0,0.3,0.3]+2(1-\hat f(0.3))\tilde c[0,0,0,0.3]}{5}\\\!\!\!\!\!&+&\!\!\!\!\!
\frac{3(1-\hat f(0)\tilde c[0,0,0.3,0.3]}{5}
=(1.99, 2.66, 0.25)
\\
\hat d_{3}\!\!\!\!\!&=&\!\!\!\!\!\hat d[0,0,0.3,0.3,0.3]=
\frac{3\hat f(0.3)\tilde d[0,0,0.3,0.3]}{5}\\\!\!\!\!\!&+&\!\!\!\!\!
\frac{2\hat f(0)\tilde 
d[0,0.3,0.3,0.3]+3(1-\hat f(0.3))\tilde c[0,0,0.3,0.3]}{5}\\\!\!\!\!\!&+&\!\!\!\!\!
\frac{2(1-\hat f(0)\tilde 
c[0,0.3,0.3,0.3]}{5}
=(2.50, 2.86, 0.40)
\\
\hat d_{4}\!\!\!\!\!&=&\!\!\!\!\!\hat d[0,0.3,0.3,0.3,0.7]=
\frac{\hat f(0.7)\tilde d[0,0.3,0.3,0.3]}{5}\\\!\!\!\!\!&+&\!\!\!\!\!
\frac{3\hat f(0.3)\tilde d[0,0.3,0.3,0.7]+
\hat f(0)\tilde d[0.3,0.3,0.3,0.7]}{5}\\&+&\frac{
(1-\hat f(0.7))\tilde c[0,0.3,0.3,0.3]}{5}\\\!\!\!\!\!&+&\!\!\!\!\!
\frac{3(1-\hat f(0.3))\tilde c[0,0.3,0.3,0.7]}{5}\\\!\!\!\!\!&+&\!\!\!\!\!
\frac{
(1-\hat f(0))\tilde c[0.3,0.3,0.3,0.7]}{5}
=(3.29, 2.52, 0.75)
\\
\hat d_{5}\!\!\!\!\!&=&\!\!\!\!\!\hat d[0.3,0.3,0.3,0.7,0.7]=
\frac{2\hat f(0.7)\tilde d[0.3,0.3,0.3,0.7]}{5}\\\!\!\!\!\!&+&\!\!\!\!\!
\frac{3\hat f(0.3)\tilde 
d[0.3,0.3,0.7,0.7]}{5}\\&+&\frac{
2(1-\hat f(0.7))\tilde c[0.3,0.3,0.3,0.7]}{5}\\\!\!\!\!\!&+&\!\!\!\!\!
\frac{3(1-\hat f(0.3))\tilde 
c[0.3,0.3,0.7,0.7]}{5}
\\\!\!\!\!\!&+&\!\!\!\!\!
(3.65, 1.64, 1.09)
\end{eqnarray*}
\begin{eqnarray*}
\hat d_{6}\!\!\!\!\!&=&\!\!\!\!\!\hat d[0.3,0.3,0.7,0.7,0.7]=
\frac{3\hat f(0.7)\tilde d[0.3,0.3,0.7,0.7]}{5}\\\!\!\!\!\!&+&\!\!\!\!\!
\frac{2\hat f(0.3)\tilde 
d[0.3,0.7,0.7,0.7]}{5}\\&+&\frac{
3(1-\hat f(0.7))\tilde c[0.3,0.3,0.7,0.7]}{5}\\\!\!\!\!\!&+&\!\!\!\!\!
\frac{2(1-\hat f(0.3))\tilde 
c[0.3,0.7,0.7,0.7]}{5}
\\\!\!\!\!\!&+&\!\!\!\!\!(3.83, 1.15, 1.30)
\\
\hat d_{7}\!\!\!\!\!&=&\!\!\!\!\!\hat d[0.3,0.7,0.7,0.7,1]=
\frac{\hat f(1)\tilde d[0.3,0.7,0.7,0.7]}{5}\\\!\!\!\!\!&+&\!\!\!\!\!
\frac{3\hat f(0.7)\tilde d[0.3,0.7,0.7,1]+
\hat f(0.3)\tilde d[0.7,0.7,0.7,1]}{5}\\&+&\frac{
(1-\hat f(1))\tilde c[0.3,0.7,0.7,0.7]}{5}\\\!\!\!\!\!&+&\!\!\!\!\!
\frac{3(1-\hat f(0.7))\tilde c[0.3,0.7,0.7,1]}{5}\\\!\!\!\!\!&+&\!\!\!\!\!
\frac{
(1-\hat f(0.3))\tilde c[0.7,0.7,0.7,1]}{5}
=(4.25, 0.62, 1.70)
\\
\hat d_{8}\!\!\!\!\!&=&\!\!\!\!\!\hat d[0.7,0.7,0.7,1,1]=
\frac{2\hat f(1)\tilde d[0.7,0.7,0.7,1]}{5}\\\!\!\!\!\!&+&\!\!\!\!\!
\frac{3\hat f(0.7)\tilde d[0.7,0.7,1,1]+2(1-\hat f(1))\tilde c[0.7,0.7,0.7,1]}{5}\\\!\!\!\!\!&+&\!\!\!\!\!
\frac{3(1-\hat f(0.7))\tilde 
c[0.7,0.7,1,1]}{5}=(5.18, 1.24, 2.15)
\\
\hat d_{9}\!\!\!\!\!&=&\!\!\!\!\!\hat d[0.7,0.7,1,1,1]=
\frac{3\hat f(1)\tilde d[0.7,0.7,1,1]}{5}\\\!\!\!\!\!&+&\!\!\!\!\!
\frac{2\hat f(0.7)\tilde d[0.7,1,1,1]+3(1-\hat f(1))\tilde c[0.7,0.7,1,1]
}{5}\\\!\!\!\!\!&+&\!\!\!\!\!
\frac{2(1-\hat f(0.7))\tilde c[0.7,1,1,1]}{5}
=(5.77, 1.30, 2.47)
\\
\hat d_{10}\!\!\!\!\!&=&\!\!\!\!\!\hat d[0.7,1,1,1,1]=
\frac{4\hat f(1)\tilde d[0.7,1,1,1]}{5}\\\!\!\!\!\!&+&\!\!\!\!\!
\frac{\hat f(0.7)\tilde d[1,1,1,1]+4(1-\hat 
f(1))\tilde c[0.7,1,1,1]}{5}\\\!\!\!\!\!&+&\!\!\!\!\!
\frac{(1-\hat f(0.7))\tilde c[1,1,1,1]}{5}
\\\!\!\!\!\!&=&\!\!\!\!\!(6.67, 0.72, 3.03)
\\
\hat d_{11}\!\!\!\!\!&=&\!\!\!\!\!\hat d[1,1,1,1,1]=\hat f(1)\tilde d[1,1,1,1]
\\\!\!\!\!\!&+&\!\!\!\!\!
(1-\hat f(1))
\tilde c[1,1,1,1]=\tilde d_{11}=(8, -1, 4).
\end{eqnarray*}

\begin{figure}[h]\begin{center}
\includegraphics[height=0.2\textheight]{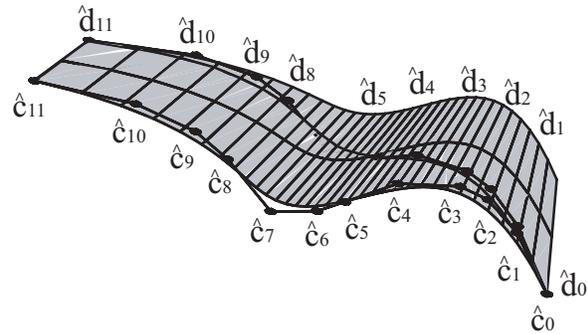}
\caption{Restriction to a triangular patch of the developable surface 
patch in Figure~\ref{tcuartica}}
\label{quintica}\end{center}
\end{figure}

The triangular B-spline net for the surface patch which satisfies the 
requirements of the example is shown in Figure~\ref{quintica}.
We check that in fact the velocity of the boundary curve $\hat d(u)$ 
of degree $n=5$ is as prescribed,
\begin{eqnarray*}\hat d'(0)&=&n\frac{\hat d_{1}-\hat d_{0}}{\hat u_{n}-\hat 
u_{n-1}}=\frac{5}{0.3}(1.20, 1.83, 0.12)\\&=&(20.00,30.50,2.00).
\end{eqnarray*}

\section{Conclusions}
\label{conclude}
We have made use of a procedure of degree elevation for obtaining
spline developable surfaces from which we know the segments of the
first and last rulings and one of the curves of the boundary.  It
consists of first solving the problem with free endpoints of the
rulings and then moving the resulting boundary curve along the rulings
to match the endpoints and increase the degree of the curves by one.
This solution is also used to solve the problem of finding a
triangular spline developable patch from which we know the last 
ruling, one of the curves of the boundary and the initial velocity of 
the other curve.

\bibliographystyle{jzus-c}
\bibliography{cagd}

\newpage

\appendix 

\section{Auxiliary points for $c(u)$\label{append}}

We perform here calculations of auxiliary points for the curve $c(u)$ 
over the list of knots $\{0,0,0,0.3,0.7,1,1,1\}$ which are needed for 
Example~\ref{spline4}, taking into account that blossoms are multiaffine 
Eq.~(\ref{maffine}):
\begin{eqnarray*}
c[0,0,0]\!\!\!\!\!&=&\!\!\!\!\!c_{0}=C_{0}=(0, 0, 0)
\\
c[0,0,0.3]\!\!\!\!\!&=&\!\!\!\!\!c_{1}=C_{1}=(2, 3, 0)
\\
c[0,0.3,0.3]\!\!\!\!\!&=&\!\!\!\!\!C_{2}=\frac{0.7-0.3}{0.7-0}c[0,0,0.3]\\&+&
\frac{0.3-0}{0.7-0}c[0,0.7,0.3]=
\frac{0.4c_{1}+0.3c_{2}}{0.7}\\\!\!\!\!\!&=&\!\!\!\!\!(2.86,3,0)
\\
c[0,0.3,0.7]\!\!\!\!\!&=&\!\!\!\!\!c_{2}=(4, 3, 0)
\\
c[0.3,0.3,0.3]\!\!\!\!\!&=&\!\!\!\!\!C_{3}=\frac{0.7-0.3}{0.7-0}c[0,0.3,0.3]\\&+&
\frac{0.3-0}{0.7-0}c[0.3,0.3,0.7]=
\frac{0.4C_{2}+0.3C_{4}}{0.7}\\\!\!\!\!\!&=&\!\!\!\!\!(3.48, 2.61, 0)
\\
c[0.3,0.3,0.7]\!\!\!\!\!&=&\!\!\!\!\!C_{4}=\frac{1-0.3}{1-0}c[0,0.3,0.7]\\&+&
\frac{0.3-0}{1-0}c[1,0.3,0.7]\\\!\!\!\!\!&=&\!\!\!\!\!
0.7c_{2}+0.3c_{3}=(4.3, 2.1, 0)
\\
c[0.3,0.7,0.7]\!\!\!\!\!&=&\!\!\!\!\!C_{5}=\frac{1-0.7}{1-0}c[0,0.3,0.7]\\&+&
\frac{0.7-0}{1-0}c[1,0.3,0.7]=
0.3c_{2}+0.7c_{3}\\\!\!\!\!\!&=&\!\!\!\!\!(4.7, 0.9, 0)
\\
c[0.3,0.7,1]\!\!\!\!\!&=&\!\!\!\!\!c_{3}=(5, 0, 0)
\\
c[0.7,0.7,0.7]\!\!\!\!\!&=&\!\!\!\!\!C_{6}=\frac{1-0.7}{1-0.3}c[0.3,0.7,0.7]\\&+&
\frac{0.7-0.3}{1-0.3}c[0.7,0.7,1]=\frac{0.3C_{5}+0.4C_{7}}{0.7}
\\\!\!\!\!\!&=&\!\!\!\!\!(5.52,1.04,0.33)
\\
c[0.7,0.7,1]\!\!\!\!\!&=&\!\!\!\!\!C_{7}=\frac{1-0.7}{1-0.3}c[0.3,0.7,1]\\&+&
\frac{0.7-0.3}{1-0.3}c[1,0.7,1]=
\frac{0.3c_{3}+0.4c_{4}}{0.7}\\\!\!\!\!\!&=&\!\!\!\!\!(6.14, 1.14, 0.57)
\\
c[0.7,1,1]\!\!\!\!\!&=&\!\!\!\!\!c_{4}=C_{8}=(7,2,1)
\\
c[1,1,1]\!\!\!\!\!&=&\!\!\!\!\!c_{5}=C_{9}=(9,-1,3).
\end{eqnarray*}

And similarly for $d(u)$,
\begin{eqnarray*}
d[0,0,0]\!\!\!\!\!&=&\!\!\!\!\!d_{0}=D_{0}=(0, 0, 2)
\\
d[0,0,0.3]\!\!\!\!\!&=&\!\!\!\!\!d_{1}=D_{1}=(1.56, 2.34, 2.08)
\\
d[0,0.3,0.3]\!\!\!\!\!&=&\!\!\!\!\!D_{2}=\frac{0.7-0.3}{0.7-0}d[0,0,0.3]
\\\!\!\!\!\!&+&\!\!\!\!\!
\frac{0.3-0}{0.7-0}d[0,0.7,0.3]=
\frac{0.4d_{1}+0.3d_{2}}{0.7}\\\!\!\!\!\!&=&\!\!\!\!\!(2.21, 2.32, 2.15)
\\
d[0.3,0.3,0.3]\!\!\!\!\!&=&\!\!\!\!\!D_{3}=\frac{0.7-0.3}{0.7-0}d[0,0.3,0.3]
\\\!\!\!\!\!&+&\!\!\!\!\!
\frac{0.3-0}{0.7-0}d[0.3,0.3,0.7]=
\frac{0.4D_{2}+0.3D_{4}}{0.7}\\\!\!\!\!\!&=&\!\!\!\!\!(2.67, 1.99, 2.24)
\\
d[0,0.3,0.7]\!\!\!\!\!&=&\!\!\!\!\!d_{2}=(3.09, 2.29, 2.26)
\\
d[0.3,0.3,0.7]\!\!\!\!\!&=&\!\!\!\!\!D_{4}=\frac{1-0.3}{1-0}d[0,0.3,0.7]
\\\!\!\!\!\!&+&\!\!\!\!\!
\frac{0.3-0}{1-0}d[1,0.3,0.7]=
0.7d_{2}+0.3d_{3}\\\!\!\!\!\!&=&\!\!\!\!\!(3.29, 1.56, 2.35)
\\
d[0.3,0.7,0.7]\!\!\!\!\!&=&\!\!\!\!\!D_{5}=\frac{1-0.7}{1-0}d[0,0.3,0.7]
\\\!\!\!\!\!&+&\!\!\!\!\!
\frac{0.7-0}{1-0}d[1,0.3,0.7]=
0.3d_{2}+0.7d_{3}\\\!\!\!\!\!&=&\!\!\!\!\!(3.55, 0.58, 2.46)
\\
d[0.3,0.7,1]\!\!\!\!\!&=&\!\!\!\!\!d_{3}=(3.75,-0.15,2.55)
\\
d[0.7,0.7,0.7]\!\!\!\!\!&=&\!\!\!\!\!D_{6}=\frac{1-0.7}{1-0.3}d[0.3,0.7,0.7]
\\\!\!\!\!\!&+&\!\!\!\!\!
\frac{0.7-0.3}{1-0.3}d[0.7,0.7,1]=\frac{0.3D_{5}+0.4D_{7}}{0.7}
\\\!\!\!\!\!&=&\!\!\!\!\!(4.15, 0.68, 2.84)
\\
d[0.7,0.7,1]\!\!\!\!\!&=&\!\!\!\!\!D_{7}=\frac{1-0.7}{1-0.3}d[0.3,0.7,1]
\\\!\!\!\!\!&+&\!\!\!\!\!
\frac{0.7-0.3}{1-0.3}d[1,0.7,1]=
\frac{0.3d_{3}+0.4d_{4}}{0.7}\\\!\!\!\!\!&=&\!\!\!\!\!(4.59, 0.75, 3.12)
\\
d[0.7,1,1]\!\!\!\!\!&=&\!\!\!\!\!d_{4}=D_{8}=(5.22, 1.42, 3.55)
\\
d[1,1,1]\!\!\!\!\!&=&\!\!\!\!\!d_{5}=D_{9}=(6.76, -1.00, 5.24).
\end{eqnarray*}

\section{Auxiliary points for $d(u)$\label{tappend}}

We compute here auxiliary points for the curve $d(u)$ 
over the list of knots $\{0,0,0,0.3,0.7,1,1,1\}$ which are needed for 
Example~\ref{splinet}, using the property of multiaffinity 
(Eq.~\ref{maffine}) for blossoms:
\begin{eqnarray*}
d[0,0,0]\!\!\!\!\!&=&\!\!\!\!\!d_{0}=(0, 0.5, 2)
\\
d[0,0,0.3]\!\!\!\!\!&=&\!\!\!\!\!d_{1}=(1.21, 2.39, 2.31)
\end{eqnarray*}
\begin{eqnarray*}
d[0,0.3,0.3]\!\!\!\!\!&=&\!\!\!\!\!\frac{0.7-0.3}{0.7-0}d[0,0,0.3]
\\\!\!\!\!\!&+&\!\!\!\!\!
\frac{0.3-0}{0.7-0}d[0,0.7,0.3]=
\frac{0.4d_{1}+0.3d_{2}}{0.7}\\\!\!\!\!\!&=&\!\!\!\!\!(1.61, 2.30, 2.67)
\\
d[0,0.3,0.7]\!\!\!\!\!&=&\!\!\!\!\!d_{2}=(2.13, 2.17, 3.16)
\\
d[0.3,0.3,0.7]\!\!\!\!\!&=&\!\!\!\!\!\frac{1-0.3}{1-0}d[0,0.3,0.7]
\\\!\!\!\!\!&+&\!\!\!\!\!
\frac{0.3-0}{1-0}d[1,0.3,0.7]=
0.7d_{2}+0.3d_{3}\\\!\!\!\!\!&=&\!\!\!\!\!(2.02, 1.50, 3.65)
\\
d[0.3,0.7,0.7]\!\!\!\!\!&=&\!\!\!\!\!\frac{1-0.7}{1-0}d[0,0.3,0.7]
\\\!\!\!\!\!&+&\!\!\!\!\!
\frac{0.7-0}{1-0}d[1,0.3,0.7]=
0.3d_{2}+0.7d_{3}\\\!\!\!\!\!&=&\!\!\!\!\!(1.88, 0.60, 4.31)
\\
d[0.3,0.7,1]\!\!\!\!\!&=&\!\!\!\!\!d_{3}=(1.77, -0.07, 4.80)
\\
d[0.7,0.7,1]\!\!\!\!\!&=&\!\!\!\!\!\frac{1-0.7}{1-0.3}d[0.3,0.7,1]
\\\!\!\!\!\!&+&\!\!\!\!\!
\frac{0.7-0.3}{1-0.3}d[1,0.7,1]=
\frac{0.3d_{3}+0.4d_{4}}{0.7}\\\!\!\!\!\!&=&\!\!\!\!\!(1.94, 0.67, 6.04)
\\
d[0.7,1,1]\!\!\!\!\!&=&\!\!\!\!\!d_{4}=(2.07, 1.22, 6.97)
\\
d[1,1,1]\!\!\!\!\!&=&\!\!\!\!\!d_{5}=(2.92, -1.00, 9.08).
\end{eqnarray*}

\section{Auxiliary points for $\tilde c(u)$\label{ttappend}}

Finally we calculate the auxiliary points which are necessary to 
formally raise the degree of the curve $\tilde c(u)$ with list of 
knots $\{0,0,0,0,0.3,0.3,0.7,0.7,1,1,1,1\}$ from four to five using the property 
of multiaffinity (Eq.~\ref{maffine}) for blossoms:
\begin{eqnarray*}
\tilde c[0,0,0,0]\!\!\!\!\!&=&\!\!\!\!\!\tilde c_{0}=(0, 0, 0)
\\
\tilde c[0,0,0,0.3]\!\!\!\!\!&=&\!\!\!\!\!\tilde c_{1}=(1.50, 2.25, 0.00)
\\
\tilde c[0,0,0.3,0.3]\!\!\!\!\!&=&\!\!\!\!\!\tilde c_{2}=(2.43, 3.00, 0.00)
\\
\tilde c[0,0.3,0.3,0.3]\!\!\!\!\!&=&\!\!\!\!\!\frac{0.7-0.3}{0.7-0}\tilde c[0,0,0.3,0.3]
\\\!\!\!\!\!&+&\!\!\!\!\!\frac{0.3-0}{0.7-0}\tilde 
c[0,0.7,0.3,0.3]\\\!\!\!\!\!&=&\!\!\!\!\!
\frac{0.4\tilde c_{2}+0.3\tilde c_{3}}{0.7}=(3.01,2.90, 0.00)
\\
\tilde c[0,0.3,0.3,0.7]\!\!\!\!\!&=&\!\!\!\!\!\tilde c_{3}=(3.79, 2.78, 0.00)
\\
\tilde c[0.3,0.3,0.3,0.7]\!\!\!\!\!&=&\!\!\!\!\!\frac{0.7-0.3}{0.7-0}\tilde c[0,0.3,0.3,0.7]
\\\!\!\!\!\!&+&\!\!\!\!\!
\frac{0.3-0}{0.7-0}\tilde c[0.7,0.3,0.3,0.7]\\\!\!\!\!\!&=&\!\!\!\!\!
\frac{0.4\tilde c_{3}+0.3\tilde c_{4}}{0.7}=(4.09, 2.23,0.00)
\end{eqnarray*}
\begin{eqnarray*}
\tilde c[0.3,0.3,0.7,0.7]\!\!\!\!\!&=&\!\!\!\!\!\tilde c_{4}=(4.50, 1.50,0.00)
\\
\tilde c[0.3,0.7,0.7,0.7]\!\!\!\!\!&=&\!\!\!\!\!\frac{1-0.7}{1-0.3}\tilde c[0.3,0.3,0.7,0.7]
\\\!\!\!\!\!&+&\!\!\!\!\!\frac{0.7-0.3}{1-0.3}\tilde 
c[1,0.3,0.7,0.7]\\\!\!\!\!\!&=&\!\!\!\!\!
\frac{0.3\tilde c_{4}+0.4\tilde c_{5}}{0.7}=(4.91, 0.93, 0.08)
\\
\tilde c[0.3,0.7,0.7,1]\!\!\!\!\!&=&\!\!\!\!\!\tilde c_{5}=(5.21, 0.51, 0.14)
\\
\tilde c[0.7,0.7,0.7,1]\!\!\!\!\!&=&\!\!\!\!\!\frac{1-0.7}{1-0.3}\tilde c[0.3,0.7,0.7,1]
\\\!\!\!\!\!&+&\!\!\!\!\!
\frac{0.7-0.3}{1-0.3}\tilde c[1,0.7,0.7,1]\\\!\!\!\!\!&=&\!\!\!\!\!
\frac{0.3\tilde c_{5}+0.4\tilde c_{6}}{0.7}=(5.99, 1.12, 0.51)
\\
\tilde c[0.7,0.7,1,1]\!\!\!\!\!&=&\!\!\!\!\!\tilde c_{6}=(6.57, 1.57, 0.79)
\\
\tilde c[0.7,1,1,1]\!\!\!\!\!&=&\!\!\!\!\!\tilde c_{7}=(7.50, 1.25, 1.50)
\\
\tilde c[1,1,1,1]\!\!\!\!\!&=&\!\!\!\!\!\tilde c_{8}=(9,-1,3).
\end{eqnarray*}

And similarly for $\tilde d(u)$,
\begin{eqnarray*}
\tilde d[0,0,0,0]\!\!\!\!\!&=&\!\!\!\!\!\tilde d_{0}=(0, 0.5, 2)
\\
\tilde d[0,0,0,0.3]\!\!\!\!\!&=&\!\!\!\!\!\tilde d_{1}=(0.91, 1.89, 2.11)
\\
\tilde d[0,0,0.3,0.3]\!\!\!\!\!&=&\!\!\!\!\!\tilde d_{2}=(1.51, 2.42, 2.20)
\\
\tilde d[0,0.3,0.3,0.3]\!\!\!\!\!&=&\!\!\!\!\!\frac{0.7-0.3}{0.7-0}\tilde d[0,0,0.3,0.3]
\\\!\!\!\!\!&+&\!\!\!\!\!
\frac{0.3-0}{0.7-0}\tilde d[0,0.7,0.3,0.3]\\\!\!\!\!\!&=&\!\!\!\!\!
\frac{0.4\tilde d_{2}+0.3\tilde d_{3}}{0.7}=(1.89, 2.35, 2.28)
\\
\tilde d[0,0.3,0.3,0.7]\!\!\!\!\!&=&\!\!\!\!\!\tilde d_{3}=(2.39, 2.24, 2.37)
\\
\tilde d[0.3,0.3,0.3,0.7]\!\!\!\!\!&=&\!\!\!\!\!\frac{0.7-0.3}{0.7-0}\tilde d[0,0.3,0.3,0.7]
\\\!\!\!\!\!&+&\!\!\!\!\!
\frac{0.3-0}{0.7-0}\tilde d[0.7,0.3,0.3,0.7]\\\!\!\!\!\!&=&\!\!\!\!\!
\frac{0.4\tilde d_{3}+0.3\tilde d_{4}}{0.7}=(2.64, 1.82, 2.37)
\\
\tilde d[0.3,0.3,0.7,0.7]\!\!\!\!\!&=&\!\!\!\!\!\tilde d_{4}=(2.97, 1.26, 2.37)
\\
\tilde d[0.3,0.7,0.7,0.7]\!\!\!\!\!&=&\!\!\!\!\!\frac{1-0.7}{1-0.3}\tilde d[0.3,0.3,0.7,0.7]
\\\!\!\!\!\!&+&\!\!\!\!\!
\frac{0.7-0.3}{1-0.3}\tilde d[1,0.3,0.7,0.7]\\\!\!\!\!\!&=&\!\!\!\!\!
\frac{0.3\tilde d_{4}+0.4\tilde d_{5}}{0.7}=(3.35, 0.77, 2.35)
\\
\tilde d[0.3,0.7,0.7,1]\!\!\!\!\!&=&\!\!\!\!\!\tilde d_{5}=(3.64, 0.39, 2.34)
\\
\tilde d[0.7,0.7,0.7,1]\!\!\!\!\!&=&\!\!\!\!\!\frac{1-0.7}{1-0.3}\tilde d[0.3,0.7,0.7,1]
\\\!\!\!\!\!&+&\!\!\!\!\!
\frac{0.7-0.3}{1-0.3}\tilde d[1,0.7,0.7,1]\\\!\!\!\!\!&=&\!\!\!\!\!
\frac{0.3\tilde d_{5}+0.4\tilde d_{6}}{0.7}=(4.53, 0.95, 2.42)
\end{eqnarray*}
\begin{eqnarray*}
\tilde d[0.7,0.7,1,1]\!\!\!\!\!&=&\!\!\!\!\!\tilde d_{6}=(5.20, 1.37, 2.48)
\\
\tilde d[0.7,1,1,1]\!\!\!\!\!&=&\!\!\!\!\!\tilde d_{7}=(6.26, 1.15, 2.87)
\\
\tilde d[1,1,1,1]\!\!\!\!\!&=&\!\!\!\!\!\tilde d_{8}=(8,-1,4).
\end{eqnarray*}







\end{document}